\documentclass[notitlepage,pre,11pt,nofootinbib,tightenlines]{revtex4}

\usepackage{amsmath}
\usepackage{amssymb,amsfonts}
\usepackage{bm}
\usepackage[mathcal]{euscript}
\usepackage{graphicx}
\usepackage{psfrag}
\usepackage{subfigure}
\usepackage{verbatim}
\usepackage{natbib}
\usepackage{enumerate}

\graphicspath{{figs/}}

\newcommand{\commentjlt}[1]{}
\newcommand{\commentgp}[1]{}
\newcommand{\jltnote}[1]{}

\newcommand{\mathnotation}[2]{\newcommand{#1}{\ensuremath{#2}}}

\renewcommand{\l}{\left}            
\renewcommand{\r}{\right}           
\mathnotation{\pd}{\partial}        
\mathnotation{\grad}{\nabla}
\mathnotation{\lapl}{\Delta}        
\mathnotation{\mlapl}{(-\Delta)}    
\mathnotation{\ee}{{\mathit{e}}}    
\mathnotation{\imi}{\mathit{i}}     
\mathnotation{\ldef}{\mathrel{\raisebox{.069ex}{:}\!\!=}}
\mathnotation{\rdef}{\mathrel{=\!\!\raisebox{.069ex}{:}}}
\mathnotation{\dint}{\,{\mathrm{d}}}

\mathnotation{\cE}{\mathcal E}      
\mathnotation{\LL}{{\mathcal L}}    
\mathnotation{\eps}{\varepsilon}    
\mathnotation{\cA}{\mathcal A}      
\mathnotation{\cK}{\mathcal K}      
\mathnotation{\kc}{k}               
\mathnotation{\kv}{{\bm{\kc}}}      
\mathnotation{\xc}{x}               
\mathnotation{\xv}{\bm{\xc}}        
\mathnotation{\uc}{u}               
\mathnotation{\uv}{\bm{\uc}}        
\mathnotation{\Uc}{U}               
\mathnotation{\Uv}{\bm{\Uc}}        
\mathnotation{\psin}{A}             
\mathnotation{\src}{s}              
\mathnotation{\rsrc}{r}             
\mathnotation{\pexp}{p}             
\mathnotation{\qexp}{q}             
\mathnotation{\acoeff}{a}           
\mathnotation{\Pe}{\mathrm{Pe}}     
\mathnotation{\Ltwo}{\mathrm{L}^2}  

\mathnotation{\cAz}{{\widetilde \cA}}
\mathnotation{\LLz}{\widetilde \LL}
\mathnotation{\thetaz}{{\widetilde \theta}}
\mathnotation{\qz}{{\widetilde \q}}

\mathnotation{\compconj}{\mathrm{c.c.}}

\newcommand{\Ltnorm}[1]{{\lVert #1 \rVert}_2}
\newcommand{\Order}[1]{\mathrm{O}(#1)}

\newcommand{\euvi}{{\hat{\bm{e}}_x}}
\newcommand{\euvj}{{\hat{\bm{e}}_y}}

\begin{document}

\title{Optimizing the Source Distribution in Fluid Mixing}
\author{Jean-Luc Thiffeault}
\email{jeanluc@imperial.ac.uk}
\author{G. A. Pavliotis}
\email{g.pavliotis@imperial.ac.uk}
\affiliation{Department of Mathematics, Imperial College
  London, SW7 2AZ, United Kingdom}

\begin{abstract}
  A passive scalar is advected by a velocity field, with a nonuniform spatial
  source that maintains concentration inhomogeneities.  For example, the
  scalar could be temperature with a source consisting of hot and cold spots,
  such that the mean temperature is constant.  Which source distributions are
  best mixed by this velocity field?  This question has a straightforward yet
  rich answer that is relevant to real mixing problems.  We use a multiscale
  measure of steady-state enhancement to mixing and optimize it by a
  variational approach.  We then solve the resulting Euler--Lagrange equation
  for a perturbed uniform flow and for simple cellular flows.  The optimal
  source distributions have many broad features that are as expected: they
  avoid stagnation points, favor regions of fast flow, and their contours are
  aligned such that the flow blows hot spots onto cold and vice versa.
  However, the detailed structure varies widely with diffusivity and other
  problem parameters.  Though these are model problems, the optimization
  procedure is simple enough to be adapted to more complex situations.
\end{abstract}


\maketitle

\section{Introduction}
\label{sec:introduction}

Consider a passive scalar advected by a velocity field, in the
presence of an inhomogeneous spatial source.  An obvious question is,
which velocity fields are best at homogenizing the concentration
field?  This is a challenging question, and here we turn it around
into a less obvious one: given a velocity field, which source
distributions are best mixed by this field?  For example, in a room
with a given airflow, where to best position heating units to achieve
a homogeneous temperature?  We will see that answering this type of
question gives considerable insight into optimal stirring flows in
general.

To carry out the optimization we need to narrow the problem further.  First,
we fix the amplitude of the velocity field, such as by specifying its total
energy (in a bounded domain) or energy density (in an unbounded domain).
Second, we restrict the source and velocity fields to be time-independent, so
that the problem is steady.  Finally, and most importantly, we need to specify
how we measure the mixing enhancement due to stirring.  Here we use a
generalization of the variance of the concentration field, which has long been
a popular measure of mixing.  The reasoning is that a velocity field that is
efficient at stirring should suppress fluctuations in the concentration field,
and the variance decreases as these fluctuations become smaller.

Previous work on optimization of mixing has focused on breaking flow
symmetries~\cite{Franjione1992,Mezic1994,Solomon2003,Grigoriew2005}, or
optimizing quantities associated with chaotic advection, such as Lyapunov
exponents or topological entropy~\cite{Sharma1997,DAlessandro1999,%
Vikhansky2002,Andrievskii2004,Balogh2005,Thiffeault2006}.  Recent work has
also involved optimizing the norm of the concentration
field~\cite{Mathew2005,Mathew2007}, in a manner similar to here.  However, all
these approaches differ from our work in that they do not involve body sources
of scalar continually replenishing the variance of the concentration field.
The optimal solutions we find are quite different, especially in that they do
not tend to lead to creation of small spatial scales in the concentration
field, but rather magnify the importance of efficient transport of temperature
from sources to sinks.

Motivated by \citet{DoeringThiffeault2006} and~\citet{Shaw2007}, we
introduce a one--parameter family of measures of mixing. The
\emph{mixing enhancement factor}~$\cE_\pexp$ is thus defined by
\begin{equation}
  \cE_\pexp \ldef \frac{\Ltnorm{\mlapl^{\pexp/2}\thetaz}}
  {\Ltnorm{\mlapl^{\pexp/2}\theta}}\,,
  \label{eq:mixeff}
\end{equation}
where~$\theta(\xv)$ is the concentration of the advected scalar,
$\Ltnorm{\cdot}$ is the~$\Ltwo$ norm, and $\lapl$ the Laplacian.
The~$\thetaz(\xv)$ in the numerator is the reference concentration obtained
for the same source distribution and diffusivity, but in the absence of
stirring.  The enhancement factor thus tells us how much better the velocity
field is at suppressing fluctuations than if we didn't stir at all.
%
%
The above definition is appropriate for the steady
advection--diffusion problem that we consider in this paper. The
relevant definition for the time-dependent problem is given in
\citet{DoeringThiffeault2006}, and involves taking the long-time
average of the numerator and denominator in~\eqref{eq:mixeff}.  As
mentioned above, for simplicity we shall restrict our study to
time-independent flows and sources, though the more general
formulation is not conceptually more complicated.  We assume without
loss of generality that the source and initial condition, and
consequently the scalar concentration, have spatial-mean zero.  For
concreteness, we will often refer to~$\theta$ as `temperature' or
`heat', and speak of `hot' and `cold' regions, but the considerations
here apply to any passive scalar.

The numerator in the definition of the enhancement~\eqref{eq:mixeff} is
designed to avoid pathological solutions, such as ones that concentrate the
source at very small scales.  Nevertheless, our definition~\eqref{eq:mixeff}
sometimes behaves counterintuitively: for instance, the enhancement can be
increased by rearranging the source on large scales to penalize diffusion.
These cases are not the most relevant ones, and do not justify introducing a
new definition of the enhancement from that used
previously~\citet{Thiffeault2004,DoeringThiffeault2006,Shaw2007}.  Moreover,
the definition~\eqref{eq:mixeff} is well-suited to optimizing the velocity
field for a fixed source distribution, which is a more important (and much
more difficult) problem for which partial results are available.  Keeping the
same definition allows comparison to these earlier results.

For~$\pexp=0$, the enhancement factor~\eqref{eq:mixeff} involves the scalar
variance and was introduced by~\citet{Thiffeault2004}.  Varying~$\pexp$
preferentially weighs the smaller ($\pexp>0$) or larger ($\pexp<0$) scales,
providing a different measure of mixing enhancement.  This measure of mixing
efficiency was used by~\citet{DoeringThiffeault2006} and~\citet{Shaw2007} in
the context of statistically-steady turbulent flows, and they found that the
scaling of the enhancement factor with the energy of the velocity field
depends strongly on the nature of the source.  For~$\pexp=-1$, the enhancement
factor is closely related to the mix-norm~\cite{Mathew2005,Mathew2007}.

In this paper we maximize~\eqref{eq:mixeff} using a variational approach
(Section~\ref{sec:optimization}).  The variation leads to an Euler--Lagrange
equation where the leading eigenvalue is the optimal enhancement factor, with
the corresponding eigenfunction giving the optimal source distribution.  Some
salient features of the optimal solutions are that (i) the optimal source
distribution avoids having hot or cold spots over stagnation points of the
flow; (ii) regions of high velocity are favored; and (iii) the hot and cold
spots are positioned such that the flow sweeps hot onto cold and vice versa.
Though most solutions have these broad features, they differ considerably in
their details and often change dramatically (but continuously) as parameters,
such as the diffusivity or the exponent~$\pexp$ in~\eqref{eq:mixeff}, are
varied.

We illustrate the range of solutions by considering first the simplest
situation, that of a uniform velocity field, as discussed
in~\citet{Plasting2006}, \citet{DoeringThiffeault2006}, and~\citet{Shaw2007}.
In that case the optimization problem can be solved analytically
(Section~\ref{sec:perturb}).  We then impose a perturbation to the uniform
flow, leading to either a shear flow or a wavy flow, and solve for the optimal
source using perturbation theory.  The resulting optimal sources favor regions
of high velocity in the shear flow, but the wavy flow leaves the optimal
solution unchanged from the uniform flow, and only decreases its efficiency.
This shows, unsurprisingly, that it is still possible to improve the
enhancement factor after optimizing the source by optimizing the velocity
field.

We then move to direct numerical solution of the optimization problem for a
simple cellular flow (Section~\ref{sec:numerics}).  The basic problem turns
out to be doubly-degenerate in that there are two independent source
distributions that give the same optimal enhancement.  This is a consequence
of a symmetry of the flow, and we verify that breaking this symmetry by adding
a small perturbation to the velocity field selects a unique optimal solution.
We also use the cellular flow to study the range of behavior as the
diffusivity and the exponent~$\pexp$ in~\eqref{eq:mixeff} are varied.  In all
these cases, the optimal source distribution converges at extreme parameter
values (large or small diffusivity or~$\pexp\rightarrow\pm\infty$).  For all
the cases, we also compare the optimized enhancement factor to two simple
reference sources, $\sin x$ and~$\cos x$, chosen to capture the spatial-phase
dependence of the optimal source distribution.  Because of the phase of the
rolls, the~$\cos x$ is much more efficient than~$\sin x$, and its enhancement
factor is very close to optimal.  This is not in itself a drawback, since it
shows that our definition of enhancement is fairly robust to changes in the
source, a desirable feature for practical implementation.  Finally, we offer
some concluding remarks in Section~\ref{sec:discussion}.

\section{The Optimization Procedure}
\label{sec:optimization}

We consider the time-independent advection--diffusion equation for a passive
scalar with concentration~$\theta(\xv)$, spatial source~$\src(\xv)$, and
diffusivity~$\kappa$,
\begin{equation}\label{e:adv_diff}
\uv(\xv) \cdot \grad \theta - \kappa \lapl \theta = \src(\xv),
\end{equation}
in $[0,L]^d$ with periodic boundary conditions. The (given) velocity
field $\uv(\xv)$ is incompressible, $\grad \cdot \uv = 0$. Both
$\uv(\xv)$ and $\src(\xv)$ are assumed to be sufficiently smooth.  We
assume that the source and initial condition have spatial mean zero,
which implies that the scalar concentration also has mean zero. We
remark that the solution $T(\xv,t)$ of the evolution problem
\begin{equation*}
\pd_t T(\xv,t) + \uv(\xv) \cdot \grad T(\xv,t) - \kappa \lapl
T(\xv,t) = \src(\xv),
\end{equation*}
converges, in the limit $t \rightarrow +\infty$, to the solution
$\theta(\xv)$ of the steady problem \eqref{e:adv_diff}, the
convergence being strong in $\Ltwo$. Hence, for steady sources and
stirrers, it is sufficient to consider the stationary problem
\eqref{e:adv_diff}.

\subsection{Optimal Mixing Enhancement}
\label{sec:optimal}

Our goal is to maximize the enhancement factor~$\cE_\pexp$ defined by
\eqref{eq:mixeff},
\begin{equation*}
  \cE_\pexp^2 \ldef \frac{\Ltnorm{\mlapl^{\pexp/2}\thetaz}^2}
  {\Ltnorm{\mlapl^{\pexp/2}\theta}^2}\,,
\end{equation*}
where $\thetaz$ solves equation \eqref{e:adv_diff} in the absence of advection,
\begin{equation}\label{e:free}
- \kappa \lapl \thetaz = \src,
\end{equation}
with periodic boundary conditions on~$[0,L]^d$.  We assume that the velocity
field is given. In maximizing the enhancement factor~$\cE_\pexp$, we fix the
$\Ltwo$ norm of the source and of the velocity field (or equivalently, the
kinetic energy of the flow).

Define the linear operators
\begin{equation*}
  \LL \ldef \uv(\xv) \cdot \grad - \kappa \lapl \qquad \text{and} \qquad \LLz
  \ldef - \kappa \lapl,
\end{equation*}
from which we can write the solution to~\eqref{e:adv_diff}
and~\eqref{e:free} as
\begin{equation*}
  \theta = \LL^{-1}\src\qquad \text{and} \qquad \thetaz = \LLz^{-1}\src\,.
\end{equation*}
We can then rewrite the enhancement factor~\eqref{eq:mixeff} as
\begin{equation}
  \cE_\pexp^2 =
  \frac{\Ltnorm{\mlapl^{\pexp/2} \LLz^{-1}\src}^2}
  {\Ltnorm{\mlapl^{\pexp/2} \LL^{-1}\src}^2}
  = \frac{\bigl\langle s\, \cAz_\pexp^{-1}\src \bigr\rangle}
  {\bigl\langle s\, \cA_\pexp^{-1}\src \bigr\rangle}\,,
  \label{eq:mixeff2}
\end{equation}
where the self-adjoint operators~$\cA_\pexp$ and~$\cAz_\pexp$ are
\begin{equation}
  \cA_\pexp \ldef \LL \mlapl^{-p} \LL^*\,,\qquad
  \cAz_\pexp \ldef \LLz \mlapl^{-p} {\LLz}^* = \kappa^2 \mlapl^{2-p}\,,
\label{eq:cA}
\end{equation}
and we have used the notation $\langle \cdot \rangle$ to denote
integration over $[0,L]^d$.  To maximize~$\cE_\pexp^2$, we compute its
variation with respect to $\src$ and set it equal to zero,
\begin{equation}
  \delta\cE_\pexp^2 = \frac{2}{\bigl\langle s\, \cA_\pexp^{-1}\src
    \bigr\rangle}\l\langle
  \l(\cAz^{-1}_\pexp s - \cE_\pexp^2\,\cA^{-1}_\pexp s\r)\delta s
  \r\rangle = 0,
  \label{eq:firstvar}
\end{equation}
which implies
\begin{equation}
  \cAz^{-1}_\pexp s = \cE_\pexp^2\,\cA^{-1}_\pexp s\,,
  \label{e:eu_la_p_GE}
\end{equation}
or
\begin{equation}
  \cA_\pexp {\cAz_\pexp}^{-1}\src = \cE_\pexp^{2}\, s.
  \label{e:eu_la_p}
\end{equation}
This is an eigenvalue problem for the operator $\cK_\pexp\ldef \cA_\pexp
{\cAz_\pexp}^{-1}$. The optimal source is given by the ground state of
the inverse of this operator, and the normalized variance is given by
the corresponding (first) eigenvalue.

The operators $\cA_\pexp^{-1}$ and $\cAz_\pexp^{-1}$ are self-adjoint
from~$\Ltwo([0,L]^d)$ to~$\Ltwo([0,L]^d)$; furthermore, they are both
positive operators in~$\Ltwo([0,L]^d)$ (restricted to functions with
mean zero).  Consequently, the generalized eigenvalue problem
\eqref{e:eu_la_p_GE} has real positive eigenvalues, and the
eigenfunctions~$\src$ and~$\src'$ corresponding to distinct
eigenvalues are orthogonal with respect to the weighted inner
product~$(\src\,,\,\src') \ldef \langle
\src\,\cAz_\pexp^{-1}\src'\rangle$.

Our goal now is to calculate the optimal source and the corresponding
mixing enhancement factor for some simple velocity fields. In particular, in
Section \ref{sec:perturb} we will consider a weakly perturbed uniform
flow in two dimensions, and in Section~\ref{sec:numerics} we will
consider cellular flows.  We will be primarily concerned
with the eigenvalue problem~\eqref{e:eu_la_p_GE} or~\eqref{e:eu_la_p}
for $\pexp=0$, i.e.
\begin{equation}
  \cA \cAz^{-1}\src = \cE^2\, \src,
  \label{e:eu_la_0}
\end{equation}
with $\cA \ldef\cA_0 = \LL \LL^*, \, \cAz \ldef \cAz_0 = \kappa^2 \lapl^2$ and
$\cE \ldef\cE_0$. Notice that the operator $\cAz^{-1}$ is a diagonal operator
in Fourier space with entries~$\kappa^2 \lvert\kv\rvert^{-4}$, where~$\kv$ is
the wavevector.  The large negative power of~$\lvert\kv\rvert$ indicates that
this operator acts as a low-pass filter, suppressing high frequencies.  We
shall return to the case~\hbox{$\pexp\ne 0$} in Section~\ref{sec:pexpdep}.

\subsection{Is the Enhancement Maximal?}
\label{sec:maximal}

Before considering specific examples in Sections~\ref{sec:perturb}
and~\ref{sec:numerics}, let us demonstrate that the optimal solution
obtained in Section~\ref{sec:optimal} is indeed a global maximum.  We
could do this using the second variation of~$\cE_\pexp^2$, but instead
we proceed more directly by expanding an arbitrary source in terms of
eigenfunctions~$s^{(i)}$ satisfying Eq.~\eqref{e:eu_la_p_GE},
\begin{equation}
  \src(\xv) = \sum_i \acoeff^{(i)} \src^{(i)}(\xv)\,,
\end{equation}
and inserting into expression~\eqref{eq:mixeff2} for the mixing enhancement
factor,
\begin{equation}
  \frac{\bigl\langle s\, \cAz_\pexp^{-1}\src \bigr\rangle}
  {\bigl\langle s\, \cA_\pexp^{-1}\src \bigr\rangle}
  = \frac{\sum_{i,j}\acoeff^{(i)}\acoeff^{(j)}
  \bigl\langle s^{(i)}\, \cAz_\pexp^{-1}\src^{(j)} \bigr\rangle}
  {\sum_{i,j}\acoeff^{(i)}\acoeff^{(j)}
    \bigl\langle s^{(i)}\, \cA_\pexp^{-1}\src^{(j)} \bigr\rangle}
  = \frac{\sum_{i,j}\acoeff^{(i)}\acoeff^{(j)}
  \bigl\langle s^{(i)}\, \cAz_\pexp^{-1}\src^{(j)} \bigr\rangle}
  {\sum_{i,j}{\cE_\pexp^{(j)}}^{-2}\acoeff^{(i)}\acoeff^{(j)}
    \bigl\langle s^{(i)}\, \cAz_\pexp^{-1}\src^{(j)} \bigr\rangle}\,,
\end{equation}
where we used the eigenfunction property.  Now use the orthogonality property
of the~$s^{(j)}$, \hbox{$\langle \src^{(i)}\,\cAz_\pexp^{-1}\src^{(j)}\rangle
= \delta_{ij}$}, and the fact that~\hbox{${\cE_\pexp} \ge
{{\cE_\pexp^{(j)}}}$}, that is the optimal enhancement factor~$\cE_\pexp$ is
the largest eigenvalue, to find
\begin{equation}
  \frac{\bigl\langle s\, \cAz_\pexp^{-1}\src \bigr\rangle}
  {\bigl\langle s\, \cA_\pexp^{-1}\src \bigr\rangle}
  = \frac{\sum_{j}(\acoeff^{(j)})^2
  \bigl\langle s^{(j)}\, \cAz_\pexp^{-1}\src^{(j)} \bigr\rangle}
  {\sum_{j}{\cE_\pexp^{(j)}}^{-2}(\acoeff^{(j)})^2
    \bigl\langle s^{(j)}\, \cAz_\pexp^{-1}\src^{(j)} \bigr\rangle}
  \le
    \frac{\sum_{j}(\acoeff^{(j)})^2
  \bigl\langle s^{(j)}\, \cAz_\pexp^{-1}\src^{(j)} \bigr\rangle}
  {\sum_{j}\cE_\pexp^{-2}(\acoeff^{(j)})^2
    \bigl\langle s^{(j)}\, \cAz_\pexp^{-1}\src^{(j)} \bigr\rangle}
  = \cE_\pexp^2\,,
\end{equation}
which proves that~$\cE_\pexp$ is indeed optimal, since the enhancement factor
of no other source can exceed it.  This is a global argument, which relies
only on the eigenvalue problem~\eqref{e:eu_la_p_GE} providing a complete set
of orthogonal eigenfunctions.

%
%
%
\section{Uniform Flow with Perturbation}
\label{sec:perturb}

As a first test case for the optimization procedure of
Section~\ref{sec:optimization}, we consider a uniform flow (constant
magnitude and direction in space).  The uniform flow illustrates a
fundamental mechanism involved in source optimization, as mentioned
in~\citet{Plasting2006} and~\citet{Shaw2007}: the velocity field
sweeps the hot source onto the cold sink, and vice versa.  More
generally, we expect that the optimal source will tend to have contours
perpendicular to the flow.  The uniform flow maximizes the mixing
enhancement factor for a one-dimensional source, at fixed kinetic energy.

To capture another important feature of optimal sources, in
Section~\ref{sec:shearwavy} we perturb the uniform flow to make either a shear
flow or a wavy flow.  The shear flow perturbation will show that optimal
sources are localized over rapid regions of the flow.  The wavy flow
perturbation will show that sometimes aligning the source contours
perpendicular to the flow is too `costly', and the optimal solution is left
unchanged from the uniform flow.  The cost incurred is due to the weighting of
the enhancement factor~\eqref{eq:mixeff} by the purely-diffusive solution: a
change in the source might make it more efficient, but it can also make the
purely-diffusive solution more efficient, yielding no net gain.

Note that in this section we will restrict our optimization to the variance,
that is with~$\pexp=0$ in the mixing enhancement~\eqref{eq:mixeff}.  We shall
return to the effect of varying~$\pexp$ in Section~\ref{sec:pexpdep}.

\subsection{Uniform Flow}
\label{sec:uniform}

For completeness, in this short section we essentially rederive the
results of~\citet{Plasting2006}, \citet{DoeringThiffeault2006},
and~\citet{Shaw2007} on the optimality of a uniform flow.  We start
with a time-independent spatially uniform flow on $[0,L]^d$,
\[
\uv(\xv) = \Uv .
\]
From definition~\eqref{eq:cA} with~$\pexp=0$, we have that
\begin{equation*}
\cA = \kappa^2 \lapl^2 - (\Uv\cdot\grad)^2\,,
\qquad
\cAz = \kappa^2 \lapl^2,
\end{equation*}
and
\begin{equation*}
\cK = \cA \cAz^{-1} = I - \kappa^{-2} (\Uv\cdot\grad)^2\lapl^{-2},
\end{equation*}
Since~$\cK$ is a differential operator with constant coefficients on a
periodic domain, its eigenfunctions are easily verified to be
\begin{equation*}
\src_\kv(\xv) = \hat{\src}_\kv \ee^{\imi \kv \cdot \xv} + \compconj\,,
\end{equation*}
where~$\kv$ is the wavevector and~$\hat{\src}_\kv$ is a normalization
constant, and the corresponding eigenvalues are
\begin{equation*}
  1 + \frac{|\Uv \cdot \kv|^2}{\kappa^2 |\kv|^4}\,.
\end{equation*}
We can maximize the above expression by choosing $\kv$ to be the shortest
allowable vector parallel to the uniform flow $\Uv$.  (Not all magnitudes and
directions are allowed for wavevectors because the domain is periodic: the
components of~$\kv$ are integer multiples of~$2\pi/L$.)

In the particular case where the uniform flow $\Uv$ is along the
$x$--axis, $\Uv = \Uc\,\euvi$, we have that
\begin{equation*}
  \cE = \sqrt{1 + \frac{\Uc^2 L^2}{4\pi^2\kappa^2}}
  \rdef \sqrt{1 + \Pe^2}
\end{equation*}
where we have defined the P\'eclet number~\hbox{$\Pe \ldef \Uc L/2\pi\kappa$}.
As expected, in the limit as $\kappa \rightarrow \infty$ ($\Pe\rightarrow 0$),
the mixing enhancement factor converges to $1$, i.e. in the large diffusivity
limit the enhancement due to a flow becomes negligible.  On the other hand, in
the limit $\kappa \rightarrow 0$ ($\Pe\rightarrow\infty$) the
enhancement factor grows like $\kappa^{-1}$.

\subsection{Shear and Wavy Flows}
\label{sec:shearwavy}

Consider now a two-dimensional uniform flow along the $x$--axis
perturbed by a weak flow,
\begin{equation}
\uv (x,y) = \Uc \euvi + \eps\, \uv_1(x,y)
- \eps^2\,\frac{\Ltnorm{\uv_1}^2}{2\Uc}\,\euvi\,,
\label{e:vel_perturb}
\end{equation}
with $\eps \ll 1$ and~$\langle\uv_1\rangle=0$. The second-order term
is included to make~$\Ltnorm{\uv} = \Uc + \Order{\eps^4}$, so that we
can compare the effect of the perturbed and unperturbed flows at equal
amplitude.  Specifically, we consider perturbations of the form
\[
\uv_1(x,y) = \uc_{1x}(y)\, \euvi + \uc_{1y}(x)\, \euvj\,,
\]
which are simple to analyze but yield important insight.  Because the base
flow is in the~$\euvi$ direction, the~$\uc_{1x}(y) \euvi$ term is a shear flow
perturbation, and the~$\uc_{1y}(x) \euvj$ term is a wavy flow perturbation.
Our goal is to obtain an asymptotic expansion for the optimal source and
enhancement factor.  For this we need to study perturbatively the eigenvalue
problem \eqref{e:eu_la_0}, which we write as
\begin{equation}
  \label{e:eu_la_perturb}
  \cA \cAz^{-1}\src = \lambda\, s\,,
\end{equation}
with $\cA = \LL \LL^*$, $\cAz = \LLz \LLz^* = \kappa^2 \lapl^2$,
and~$\lambda=\cE^2$. We have that
\begin{equation*}
  \LL = \LL_0 + \eps \LL_1 + \eps^2 \LL_2\,,
\end{equation*}
with
\begin{equation*}
  \LL_0 = \Uc \pd_x - \kappa \lapl\,, \qquad
  \LL_1 = \uv_1 \cdot \grad\,,\qquad
  \LL_2 = \uc_{2x}\pd_x\,,
\end{equation*}
where~$\uc_{2x}$ is the coefficient of the second-order term
in~\eqref{e:vel_perturb}.  Consequently,
\begin{align*}
  \cA  &= \LL_0 \LL_0^* + \eps (\LL_1 \LL_0^* - \LL_0 \LL_1) + \eps^2
  (\LL_2 \LL_0^* - \LL_0 \LL_2 - \LL_1 \LL_1) \\
  &\rdef \cA_0 + \eps \cA_1 + \eps^2 \cA_2.
\end{align*}
Note that in this section the subscripts on the~$\cA$'s correspond to
their order in~$\eps$, and not to the~$\pexp$ subscript as in
Eq.~\eqref{eq:cA} (recall that~$\pexp=0$ for the present section).  We
expand $\src$ and $\lambda$ in a power series in $\eps$,
\begin{align*}
\src &= \src_0 + \eps \src_1 + \eps^2 \src_2 + \dots, \\
\lambda &= \lambda_0 + \eps \lambda_1 + \eps^2 \lambda_2 + \dots,
\end{align*}
insert the expansions for $\cA, \, s$ and $\lambda$ in equation
\eqref{e:eu_la_perturb}, and equate like powers of $\eps$ to obtain
the sequence of equations
\begin{subequations}
\begin{gather}
\label{e:eps_0}
\cA_0 \cAz^{-1}\src_0 = \lambda_0 \src_0\,,\\
\label{e:eps_1}
\cA_1 \cAz^{-1}\src_0 + \cA_0 \cAz^{-1}\src_1 = \lambda_0 \src_1 +
\lambda_1 \src_0\,,\\
\label{e:eps_2}
\cA_2 \cAz^{-1}\src_0 + \cA_1 \cAz^{-1}\src_1 + \cA_0 \cAz^{-1}\src_2 =
\lambda_0 \src_2 + \lambda_1 \src_1 + \lambda_2 \src_0\,.
\end{gather}
\end{subequations}
Consider first equation \eqref{e:eps_0}, the unperturbed equation: from
Section~\ref{sec:uniform} we have that
\[
\src_0 = \hat{\src}_0 \ee^{\imi \kc_0 x} + \compconj\,,\qquad
\lambda_0 = 1 + \frac{\Uc^2}{\kappa^2 \kc_0^2} = 1 + \Pe^2\,.
\]
where~\hbox{$\kc_0 \ldef 2\pi/L$}.  We proceed now with
equation~\eqref{e:eps_1}: multiply by $\src_0$ and integrate over
$[0,L]^2$ to obtain
\begin{equation}\label{e:eps_11}
\langle \src_0\, \LL_1 \LL_0^* \cAz^{-1}\src_0\rangle
- \langle \src_0\, \LL_0 \LL_1 \cAz^{-1}\src_0\rangle
+ \langle \src_0\, \cA_0
\cAz^{-1}\src_1\rangle = \lambda_0 \langle \src_0\, \src_1\rangle
+ \lambda_1 \Ltnorm{\src_0}^2\,.
\end{equation}
Notice that $\cA_0$ is a differential operator with constant
coefficients, and hence it commutes with $\cAz^{-1}$. Furthermore, both
operators are self-adjoint, so that
\begin{equation}
\langle \src_0\, \cA_0 \cAz^{-1}\src_1\rangle = \langle \src_0\, \cAz^{-1}
  \cA_0
  \src_1\rangle
  =  \langle(\cA_0 \cAz^{-1}\src_0)\, \src_1\rangle
  = \lambda_0 \langle \src_0\, \src_1\rangle
  \label{e:int_parts}
\end{equation}
and equation \eqref{e:eps_11} simplifies to
\begin{equation*}
\langle \src_0\, \LL_1 \LL_0^* \cAz^{-1}\src_0\rangle - \langle \src_0\, \LL_0
 \LL_1
 \cAz^{-1}\src_0\rangle = \lambda_1 \Ltnorm{\src_0}^2
\end{equation*}
leading to
\begin{equation*}
\lambda_1 = \frac{\langle \src_0\, \LL_1 \LL_0^* \cAz^{-1}
  \src_0\rangle}{\Ltnorm{\src_0}^2}
- \frac{\langle \src_0\, \LL_0 \LL_1 \cAz^{-1}\src_0\rangle}{
  \Ltnorm{\src_0}^2}
= 0,
\end{equation*}
where the last equality follows from a straightforward calculation.  This is a
consequence of the invariance of the enhancement factor under reversal of the
pertubation, from which all odd powers of~$\eps$ in the enhancement factor
will vanish.

Now we compute $\src_1$. We set $\lambda_1 = 0$ in equation
\eqref{e:eps_1} to obtain
\begin{equation}
\label{e:s1}
\big(\lambda_0 - \cA_0 \cAz^{-1}  \big) \src_1 = \cA_1 \cAz^{-1}\src_0.
\end{equation}
For definiteness, we specify the form of the perturbation,
\[
\uc_{1x}(y) = \hat{\uc}_{1x} \ee^{\imi \kc_1 y} + \compconj
\]
Notice that we do not yet need to specify $\uc_{1y}(x)$: it only affects
the result at the next order.  We look for a solution to \eqref{e:s1}
in the form of
\begin{equation}
\src_1(x,y) = \widehat{A}_{\kc_0 \kc_1} \hat{\src}_0 \hat{\uc}_{1x} \ee^{\imi
  (\kc_0 x + \kc_1 y)} +
\widehat{B}_{\kc_0 \kc_1} \hat{\src}_0 \overline{\hat{\uc}}_{1x}
\ee^{\imi (\kc_0 x - \kc_1 y)} + \compconj,
  \label{e:s1sol}
\end{equation}
and find
\begin{equation}
\widehat{A}_{\kc_0 \kc_1} = \widehat{B}_{\kc_0 \kc_1} = \frac{ (\kc_0^2 +
  \kc_1^2)^2}{\Uc^2\, \kc_1^2\, (2 \kc_0^2 + \kc_1^2)}
  \left(2 \Uc - \imi\, \frac{\kappa \kc_1^2}{\kc_0} \right).
  \label{e:s1solAB}
\end{equation}

Let us proceed now with the calculation of $\lambda_2$. We multiply
\eqref{e:eps_2} by $\src_0$, integrate over $[0,L]^2$ and use
\eqref{e:int_parts} to obtain
\begin{equation*}
\lambda_2 = \frac{\langle \src_0\, \cA_2
  \cAz^{-1}\src_0\rangle}{\Ltnorm{\src_0}^2} + \frac{\langle \src_0\,
  \cA_1 \cAz^{-1} \src_1\rangle}{\Ltnorm{\src_0}^2}\,,
\end{equation*}
which after substituting our
solution~\eqref{e:s1sol}--\eqref{e:s1solAB} for~$\src_1$ simplifies to
\begin{equation*}
  \lambda_2 = \frac{(4\kc_0^2\,\Uc^2 + \kc_1^4\kappa^2)}
  {\Uc^2\,\kappa^2  \kc_1^2 (2\kc_0^2 + \kc_1^2)}\,\Ltnorm{\uc_{1x}}^2
  - \frac{1}{\kappa^2\,\kc_0^2}\,\Ltnorm{\uc_{1y}}^2\,.
\end{equation*}
\jltnote{Using Mathematica, I got the general form for arbitrary $\pexp$.
  Nice to see the asymptotic behavior with $\pexp$, which does show an
  exponential growth, as well as `symmetry' in $\pm\pexp$.  Also exhibits
  symmetry about~\hbox{$\pexp=1$}, which seems to be a general
  feature.  Shown in Appendix~\ref{sec:psym} for large~$\kappa$: show
  this in general!}
We are really after the enhancement factor, which is
\[
  \cE = \sqrt{\lambda_0 + \eps^2\lambda_2} + \Order{\eps^4}
  = \sqrt{\lambda_0} + \eps^2\frac{\lambda_2}{2\sqrt{\lambda_0}}
  + \Order{\eps^4}\,,
\]
so after putting it all together, we can write~$\cE$ in terms of
dimensionless quantities,
\begin{equation}
  \cE = \sqrt{1 + \Pe^2} + \eps^2\,\frac{\Pe^2}{2\sqrt{1 + \Pe^2}}
  \,\l\{\frac{(4 + \alpha^4\,\Pe^{-2})}
  {\alpha^2  (2 + \alpha^2)}\,\frac{\Ltnorm{\uc_{1x}}^2}{\Uc^2}
  - \frac{\Ltnorm{\uc_{1y}}^2}{\Uc^2}\r\} + \Order{\eps^4}\,,
  \label{e:lambda2final3}
\end{equation}
where~$\alpha\ldef\kc_1/\kc_0$ and recall that~$\Pe=\Uc/\kappa\kc_0$.

Some observations regarding~\eqref{e:lambda2final3} are in order.
First, notice that a nonzero~$\uc_{1x}$ (shear flow perturbation)
increases the enhancement factor, while a nonzero~$\uc_{1y}$ (wavy flow)
decreases it.  Second, the form of~$\uc_{1y}$ is irrelevant, and it
does not affect the first-order optimal source distribution.
\jltnote{This seems to be true to all orders!  Should be easy to
  show.}  Third, in the limit of large~$\kappa$ (small~$\Pe$),
the~$\Pe^{-2}$ term in~\eqref{e:lambda2final3}, arising
from~$\kappa^2\lapl^2$ in~$\cA_0$, invalidates the expansion.  We
discuss each type of perturbation in turn.

\begin{figure}
\psfrag{x}{$x$}
\psfrag{y}{$y$}
\psfrag{sev1}{}
\begin{center}
\subfigure[]{
\includegraphics[width=.45\textwidth]{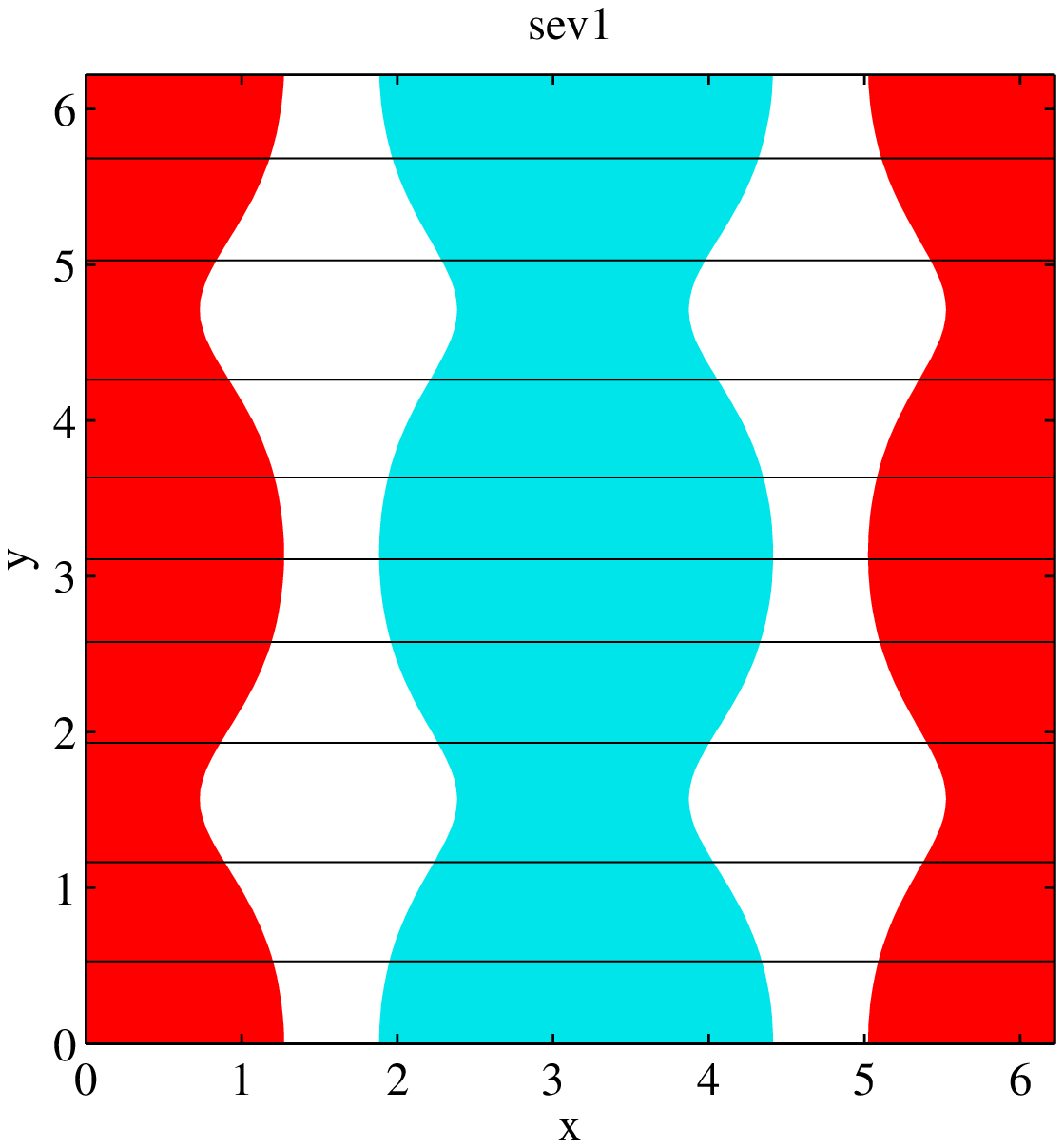}
\label{fig:srcopt_pertx}
}\hspace{1em}
\subfigure[]{
\includegraphics[width=.45\textwidth]{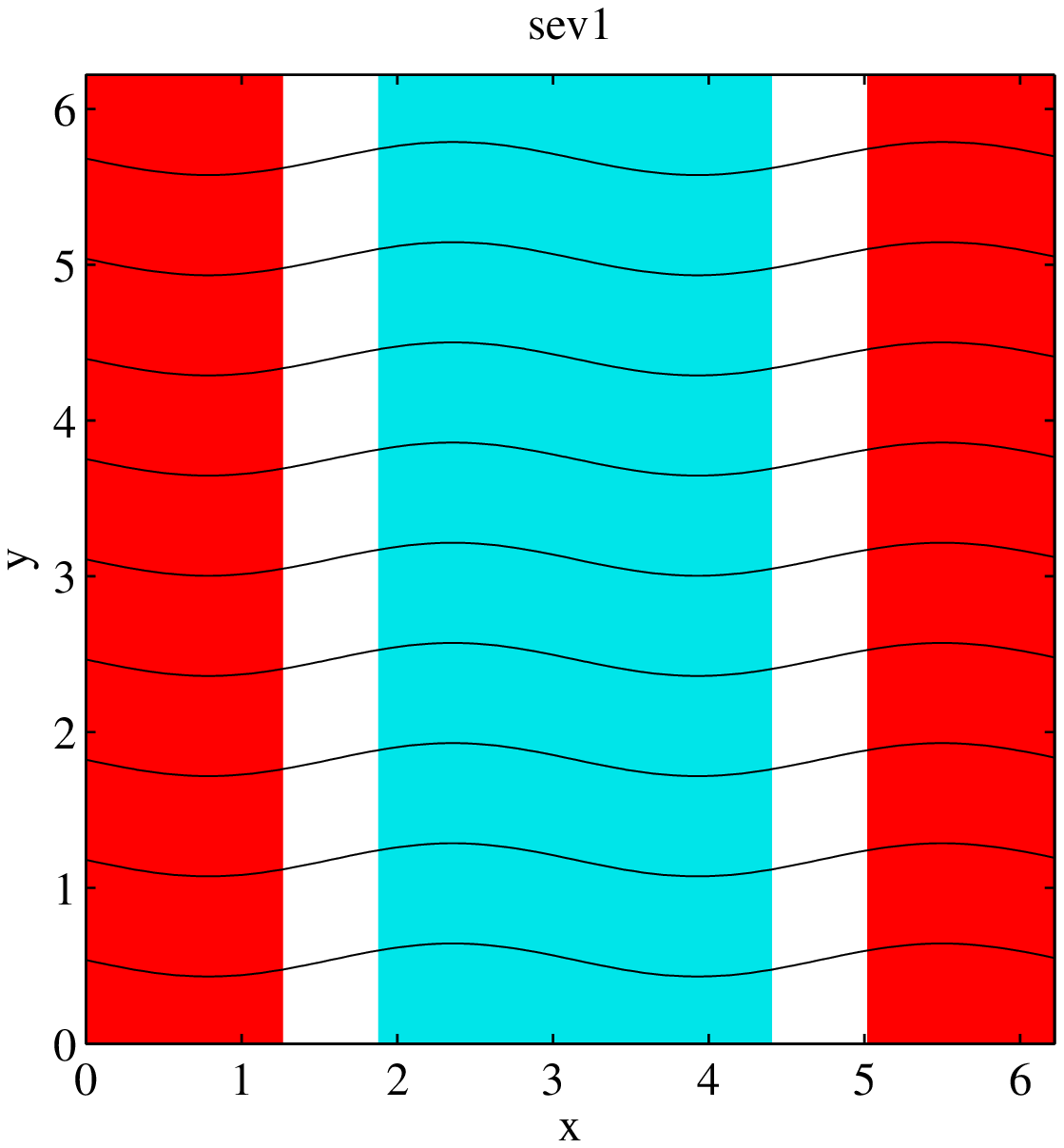}
\label{fig:srcopt_perty}
}
\end{center}
\caption{Optimal source distribution for the flow~$\euvi + \eps\,
  \uv_1(x,y) - \eps^2\,\tfrac{1}{2}\Ltnorm{\uv_1}^2\,\euvi$ with
  $\eps=0.15$, $\kappa=0.01$ and (a) $\uv_1(y) = \sqrt{2}\cos
  2y\,\euvi$; (b) $\uv_1(x) = \sqrt{2}\cos 2x\,\euvj$.  The
  horizontal phase of the solution is arbitrary.  The background
  shading shows hot (red, or dark gray) and cold (blue, or light gray)
  regions, separated by tepid regions (white).  The contour lines are
  streamlines of the flow.}
\label{fig:srcopt_pert}
\end{figure}

Figure~\ref{fig:srcopt_pertx} shows the optimal source distribution
for $U=1$, $L=2\pi$, $\kappa=0.01$ ($\Pe=100$),
$\hat{\uc}_{1x}=1/\sqrt{2}$, $\kc_1=2$, $\eps=0.15$, $\uc_{1y}=0$.
This is a shear flow, with contour lines shown in the background.  The contour
lines are closer together at the midpoint, indicating faster flow, since the
perturbation is~$\uc_{1x}(y)=\sqrt{2}\cos 2y$.  Accordingly, the optimal
source is localized at these points of faster flow.  However, there is a limit
to how localized it can get: if the source were to bunch up too much in the
faster region, the purely-diffusive solution would be more effective (i.e.,
have lower variance), \emph{reducing} the enhancement factor.  As the
perturbation gets larger, the optimal source will bunch up more in the faster
regions, since the gain becomes greater.  The enhancement factor is always
greater than in the unperturbed case, suggesting that the advantage of faster
flow regions outweighs the disadvantage of localizing the source.  Finally,
note that for large~$\alpha$ (large perturbation wavenumber~$\kc_1$), the
mixing enhancement factor becomes independent of~$\alpha$, reflecting the fact
that a fine-scale perturbation is overwhelmed by diffusion.

Conversely, Figure~\ref{fig:srcopt_perty} shows the optimal source
distribution for the same parameters as the shear flow above but with
perturbation~$\uc_{1y}=\sqrt{2}\cos 2x$, $\uc_{1x}=0$.  The streamlines in the
background indicate that this is a wavy flow, with no dependence on~$y$.  The
optimal source distribution, again shown in the background (here as~$\cos x$,
but the phase is arbitrary), is exactly the same as in the absence of
perturbation.  Intuitively, one could expect the source to be more efficient
if it tilted to present its contours perpendicular to the flow, giving a wavy
source.  This is not the case, since the purely-diffusive solution would be
more effective for a wavy source, and thus lower the enhancement factor since
it enters the numerator in the definition~\eqref{eq:mixeff}.  Localizing the
source in regions of faster flow (as for the perturbation in
Fig.~\ref{fig:srcopt_pert}) is a more important effect than aligning the
contours of the source perpendicular to the flow.

\jltnote{$\kappa$ dependence.  Independent of $\kc_1$ for large
  $\kc_1$.}

%
%

\section{Numerical Results for the Cellular Flow}
\label{sec:numerics}

In Section~\ref{sec:perturb} we derived expressions for optimal sources and
mixing enhancement factors for perturbations of a uniform flow.  More general
velocity fields require a numerical approach.  To maximize the enhancement
factor, we again have to solve the generalized eigenvalue
problem~\eqref{e:eu_la_p} for the optimal source distribution~$\src$.  For
numerical implementation, it is preferable to solve the equivalent
self-adjoint eigenvalue problem
\begin{equation}
  (\cAz^{-1/2}\cA\cAz^{-1/2})\,\rsrc = \cE^2\,\rsrc\,,
  \qquad
  \src = \cAz^{1/2}\,\rsrc\,,
  \label{eq:ge}
\end{equation}
for the eigenvector~$\rsrc$, which then yields the optimal source
distribution~$\src$.  The advantage of the form~\eqref{eq:ge} is that the
self-adjoint structure of the operator is explicit.  In practice, we
expand~$\rsrc$ as a Fourier series, so that~$\cAz$ is a diagonal matrix.  We
then solve~\eqref{eq:ge} using Matlab's \texttt{eigs} routine for sparse
matrices.  We set the box size~$L=2\pi$ throughout this section.  Note that we
shall deal only with~$\pexp=0$ (variance optimization) until
Section~\ref{sec:pexpdep}, so we leave off the~$\pexp$ subscript until then.

\subsection{Cellular Flow}
\label{sec:cellflow}

We consider the perturbed cellular flow with streamfunction
\begin{equation}
  \psi(x,y) = \psin\l(\sin x\, \sin y + \delta_1\,\sin 2x + \delta_2\,\sin
  2x\,\sin 2y\r)
  \label{eq:sf}
\end{equation}
with velocity field~\hbox{$\uv =
  (\uc_x,\uc_y)=(\pd_y\psi,-\pd_x\psi)$}, where~$\psin$ is a
normalization constant chosen to make~\hbox{$\lVert\uv\rVert_2 = 1$}.
With~$\delta_1=\delta_2=0$ (the basic cellular flow), the
operator~$\LL$ is invariant under the discrete symmetry
group~$\mathcal{G}$ generated by the transformations
\begin{subequations}
\begin{alignat}{2}
  G_1\,(x,y) &= (y\,,\,-x+\pi), \qquad && \text{rotation with vertical
    translation};
  \label{eq:G1}\\
  G_2\,(x,y) &= (x+\pi\,,\,y+\pi), \qquad && \text{diagonal translation}.
  \label{eq:G2}
\end{alignat}%
\label{eq:G}%
\end{subequations}%
The Abelian group~$\mathcal{G}$ has order~$8$ and is characterized
by~$G_1^4=G_2^2=I$, $G_1G_2=G_2G_1$.%
\footnote{$\mathcal{G}$ is the direct product of a cyclic group of
  order 4 and a cyclic group of order 2.}
For either~$\delta_1$ or~$\delta_2$
nonzero, the perturbations break the symmetry~$G_1$ of the cellular flow, and
we shall look at their effect in turn.

First we set~$\delta_1=\delta_2=0$ in~\eqref{eq:sf} and solve~\eqref{eq:ge}
with~$\kappa=0.01$.
\begin{figure}
\psfrag{x}{$x$}
\psfrag{y}{$y$}
\psfrag{sev1}{$\src^{(1)}$}
\psfrag{sev2}{$\src^{(2)}$}
\begin{center}
\subfigure{
\includegraphics[width=.45\textwidth]{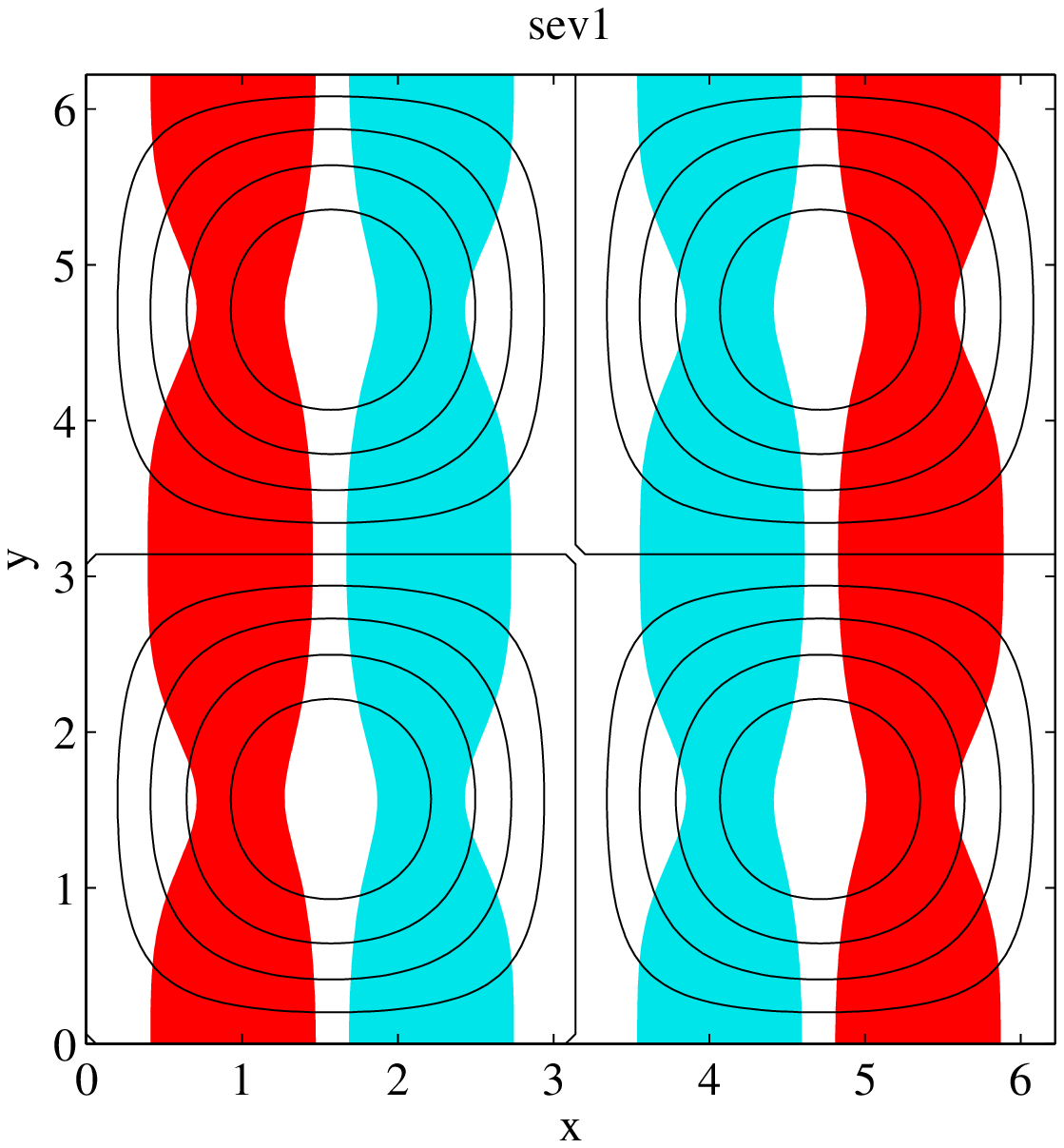}
}\hspace{1em}
\subfigure{
\includegraphics[width=.45\textwidth]{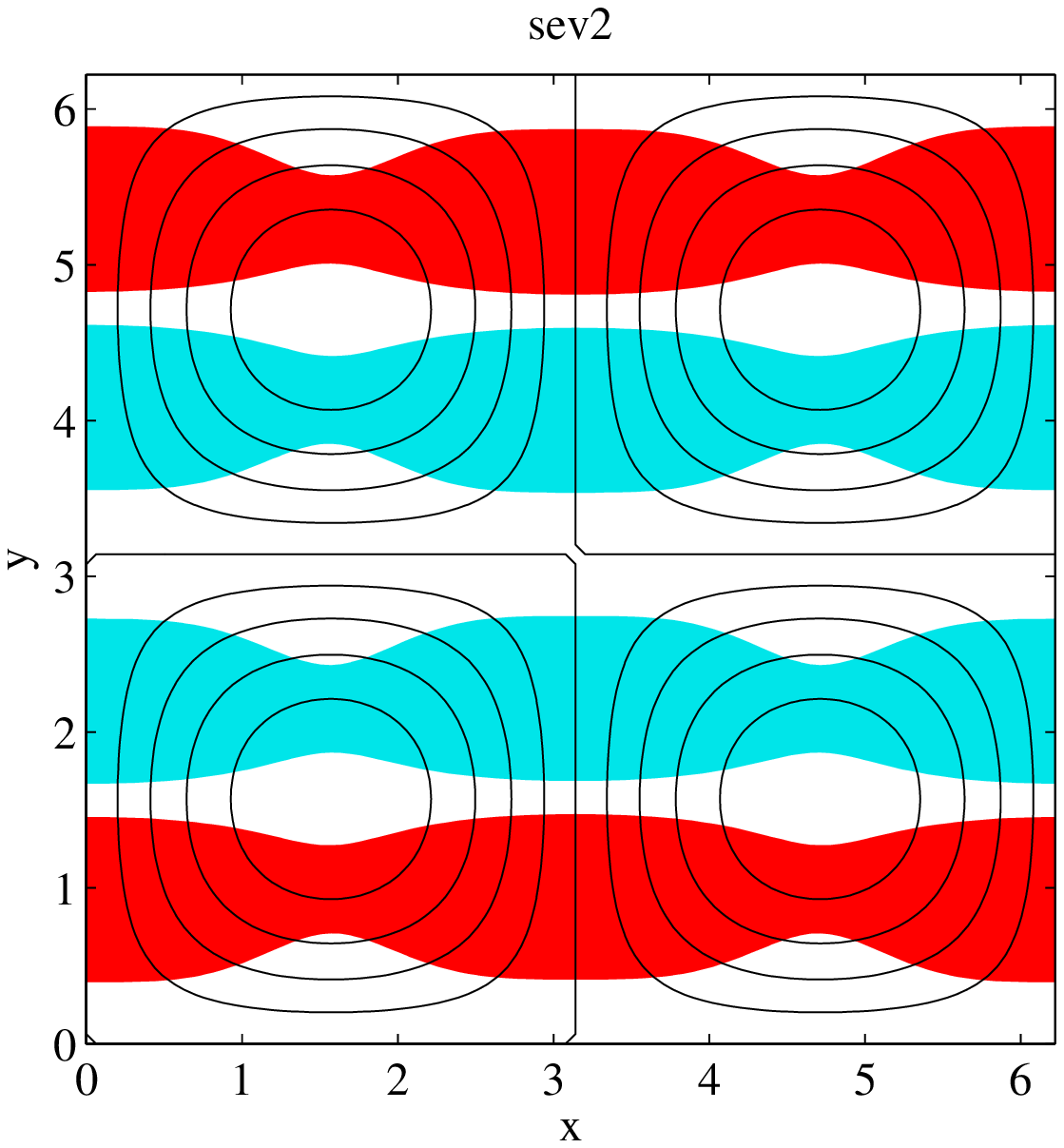}
}
\end{center}
\caption{Optimal source distributions for the pure cellular flow~\eqref{eq:sf}
  with \hbox{$\delta_1=\delta_2=0$}.  The two pictures show degenerate
  orthogonal eigenfunctions with enhancement factor~\hbox{$\cE=87.34$}.  Note
  how there is no source of heat over the stagnation points.  (See the caption
  to Fig.~\ref{fig:srcopt_pert} for a key to the background shading.)}
\label{fig:srcopt_cell}
\end{figure}
In this case, there are two independent optimal source eigenfunctions
with degenerate enhancement factor~$\cE=87.34$, shown in
Fig.~\ref{fig:srcopt_cell}.  In the foreground are contour lines of
the streamfunction.  Any superposition of the two eigenfunctions in
Fig.~\ref{fig:srcopt_cell} will give the same enhancement factor.  The
degeneracy is a consequence of the symmetry~$G_1$ of the cellular
flow: indeed, the two eigenfunctions are related to each other (up to
a sign) by the transformation~\eqref{eq:G1}.  The two eigenfunctions
also separately have the~$G_2$ symmetry.  This situation, where the
eigenfunctions corresponding to the same eigenvalue are related by
a unitary representation of the symmetry group of an operator, is
familiar from quantum mechanics~\cite{Wigner,Hamermesh}.

For comparison, the enhancement factor for the same flow but with the
reference source~$\src(\xv) = \sin x$ is~$50.01$, and with the
reference source~$\cos x$ is~$86.61$.  Hence, for~$\sin x$ the optimal
source gives a~$74.7\%$ improvement in the enhancement factor, but
only~$0.9\%$ for~$\cos x$.  It is remarkable how close~$\cos x$ comes
to the optimal enhancement factor, which shows that the optimal source is
rather `robust', so that small changes in its shape do not lead to
huge changes in the enhancement factor.  From a design standpoint, this is
highly desirable.  Table~\ref{tab:eff} summarizes the optimal
enhancement factor results for different values of the
perturbations~$\delta_1$ and~$\delta_2$.
\begin{table}
  \caption{Optimal mixing enhancement factors~$\cE_{\mathrm{optimal}}$
    with~$\pexp=0$ for the perturbed cellular flow~\eqref{eq:sf},
    with~$\kappa=0.01$.  Compare to the reference enhancements for a~$\sin x$
    and~$\cos x$ source distribution: the number in parentheses is the~$\%$
    improvement of the optimal source.  The~$\cos x$ source always does much
    better than~$\sin x$ because it straddles the rolls, whereas~$\sin x$ has
    hot and cold segregated into different rolls.}
\medskip
\begin{ruledtabular}
\begin{tabular}{cccccl}
$\delta_1$ & $\delta_2$ & $\cE_{\mathrm{optimal}}$ & $\cE_{\sin}$
($\%$) & $\cE_{\cos}$ ($\%$) & note \\
\hline
$0$ & $0$ & $87.34$ & $50.01$ ($74.7\%$) & 86.61 ($0.9\%$) &
Fig.~\ref{fig:srcopt_cell} (degenerate)\\
\hline
$0.05$ & $0$ & $87.59$ & $49.76$ ($76.0\%$) & 86.18 ($1.6\%$) &
Fig.~\ref{fig:srcopt_cell_delta1}\\
$0$ & $0.05$ & $90.10$ & $50.26$ ($79.3\%$) & 86.47 ($4.2\%$) &
Fig.~\ref{fig:srcopt_cell_delta2}\\
$0.05$ & $0.05$ & $90.19$ & $50.01$ ($80.3\%$) & 86.04 ($4.8\%$)\\
\hline
$0.2$ & $0$ & $90.43$ & $46.43$ ($94.7\%$) & 80.41 ($12.5\%$)\\
$0$ & $0.2$ & $95.34$ & $53.35$ ($78.7\%$) & 84.59 ($12.7\%$)\\
$0.2$ & $0.2$ & $97.60$ & $50.01$ ($95.2\%$) & 79.30 ($23.1\%$) &
Fig.~\ref{fig:srcopt_cell_delta12}\\
\hline
$0.5$ & $0$ & $95.99$ & $35.37$ ($171\%$) & 61.25 ($56.7\%$) &
Fig.~\ref{fig:srcopt_cell_delta1b}\\
$0$ & $0.5$ & $94.91$ & $61.25$ ($55.0\%$) & 79.06 ($20.1\%$)\\
$0.5$ & $0.5$ & $106.8$ & $50.01$ ($114\%$) & 64.55 ($65.4\%$)
\end{tabular}
\end{ruledtabular}
\label{tab:eff}
\end{table}

\begin{figure}
\psfrag{x}{$x$}
\psfrag{y}{$y$}
\psfrag{|u|}{}
\begin{center}
\includegraphics[width=.5\textwidth]{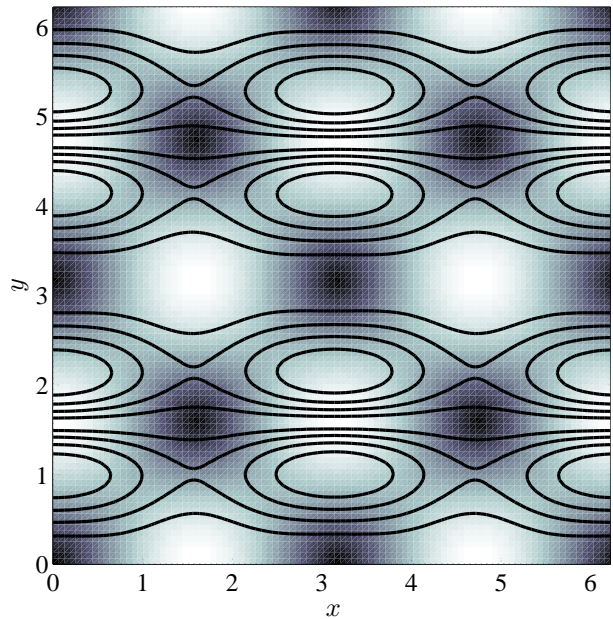}
\end{center}
\caption{The background shows the magnitude~${\lvert\uv\rvert}$ of the
  velocity field for the pure cellular flow, Eq.~\eqref{eq:sf}
  with~$\delta_1=\delta_2=0$.  The contour lines are from the optimal source
  distribution on the right in Fig.~\ref{fig:srcopt_cell}.  Note how the
  source-sink pairs (rolls in the contours) are clustered over regions of high
  speed (pale) and avoid the stagnation points (dark).}
\label{fig:srcopt_cell_umag}
\end{figure}

How to interpret the optimal source distribution in
Fig.~\ref{fig:srcopt_cell}?  The lesson learned from the uniform flow
of Section~\ref{sec:perturb} is that, ideally, the source should be such that
the velocity field advects heat from hot to cold and vice versa.  This
is clearly happening in Fig.~\ref{fig:srcopt_cell} to some extent,
since in both eigenfunctions the hot and cold spots are situated such
that the rolls easily advect heat between hot and cold.  However, this
is not the whole story: the optimal source is also distributed so as
to take advantage of regions of faster flow.  This is readily apparent
in Fig.~\ref{fig:srcopt_cell_umag}, which shows the magnitude of the
velocity field in the background,
and contours of the optimal source distribution from
Fig.~\ref{fig:srcopt_cell} (right).  The hot and cold spots (elliptic
regions) are clearly localized over the regions of rapid flow (pale
background). Some of the fast regions appear to have no hot or cold
spots, but these regions are favored in the other eigenfunction, so
the symmetry is respected.

\begin{figure}
\psfrag{x}{$x$}
\psfrag{y}{$y$}
\psfrag{sev}{}
\begin{center}
\subfigure[]{
  \includegraphics[width=.4\textwidth]{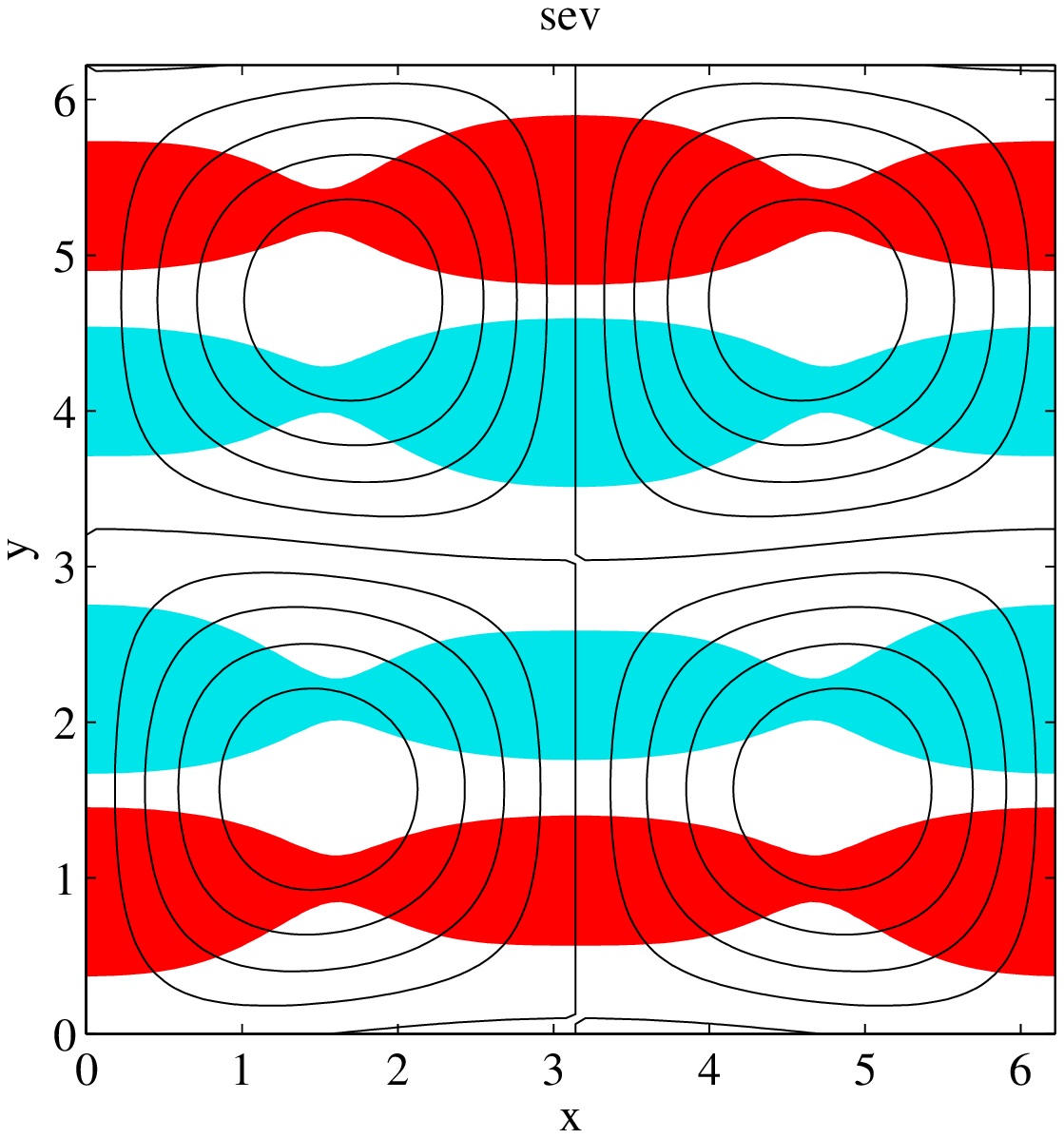}
  \label{fig:srcopt_cell_delta1}
}\hspace{1em}%
\subfigure[]{
  \includegraphics[width=.4\textwidth]{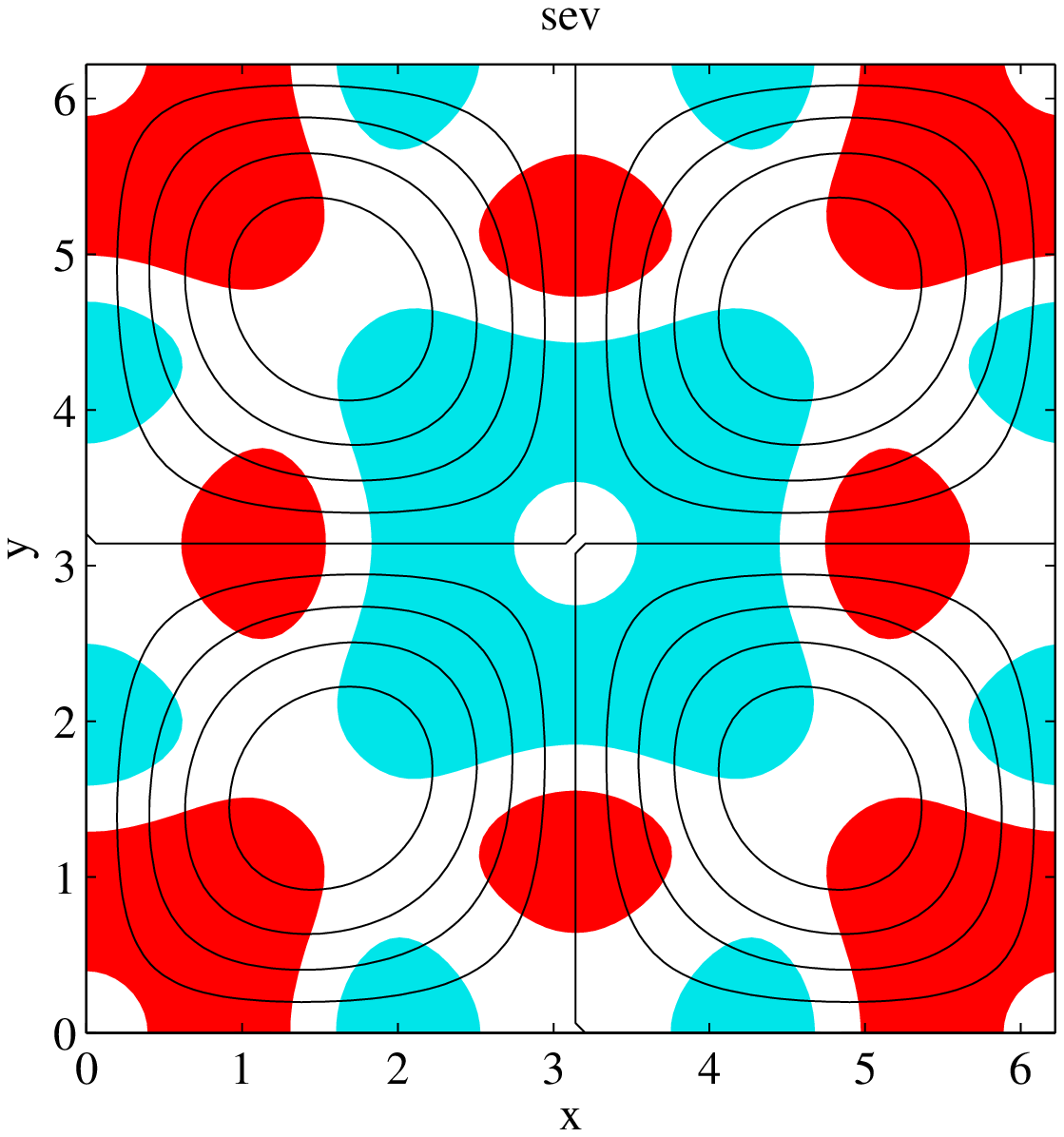}
  \label{fig:srcopt_cell_delta2}
}

\subfigure[]{
  \includegraphics[width=.4\textwidth]{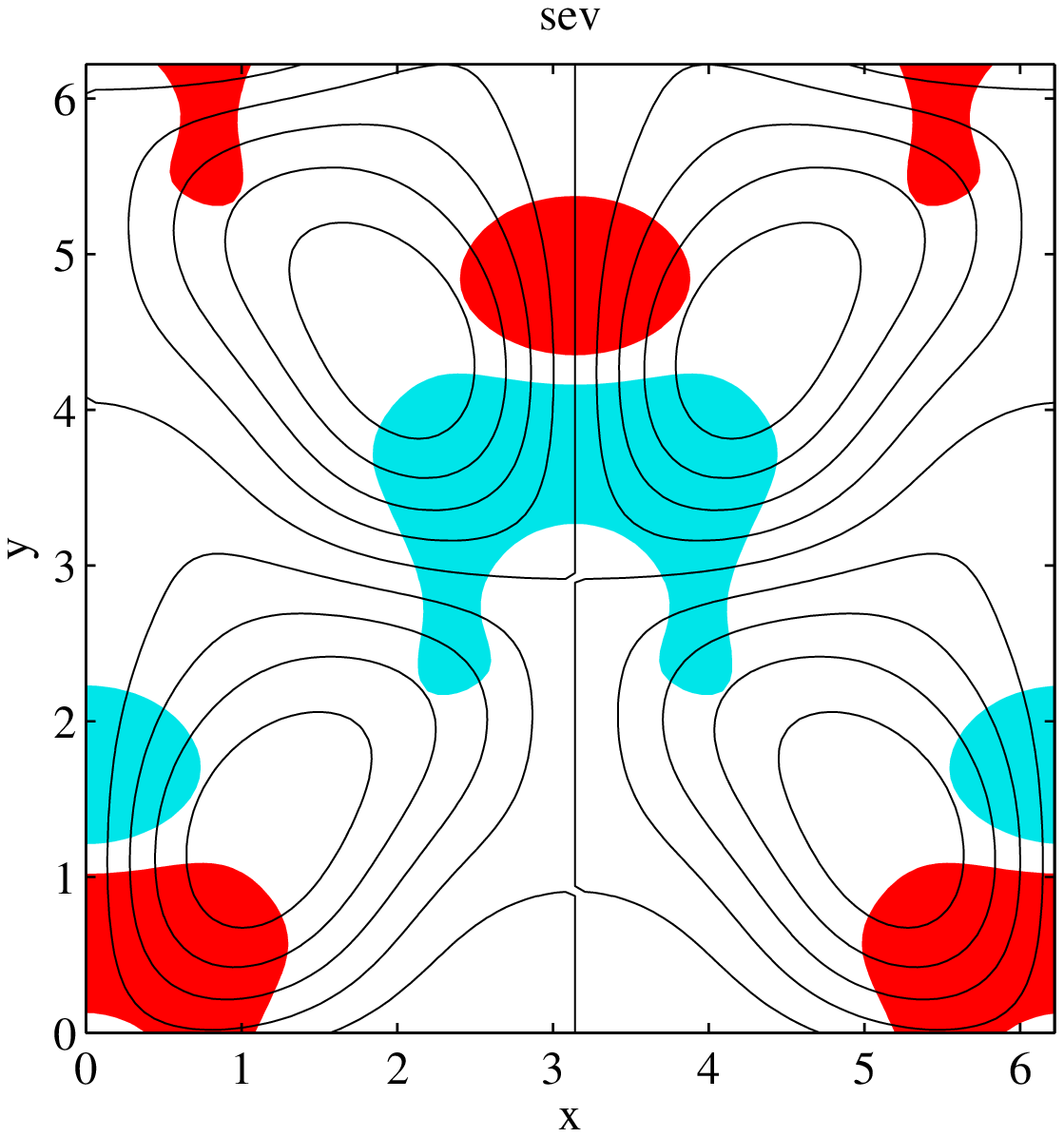}
  \label{fig:srcopt_cell_delta12}
}\hspace{1em}%
\subfigure[]{
  \includegraphics[width=.4\textwidth]{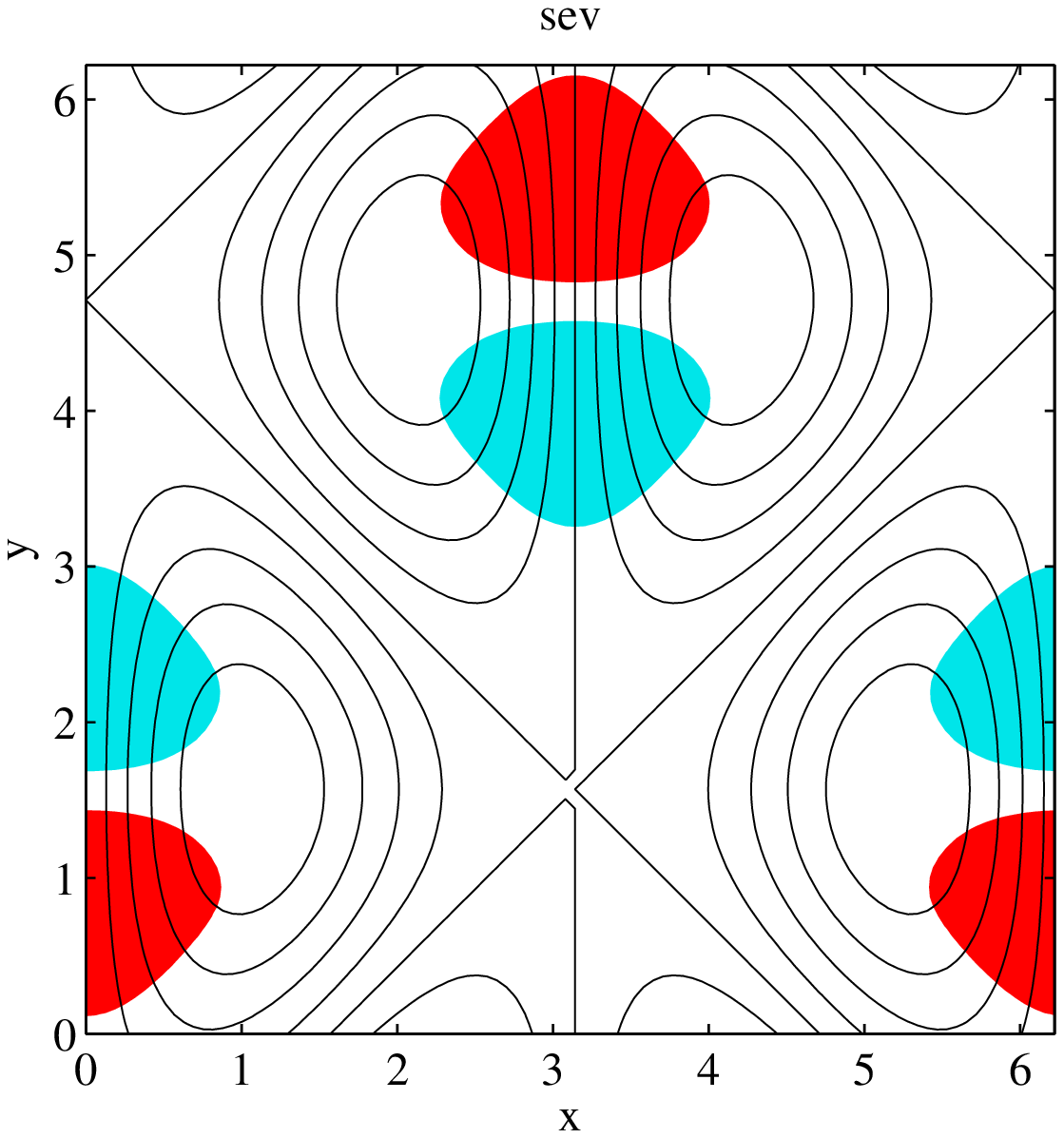}
  \label{fig:srcopt_cell_delta1b}
}
\end{center}
\caption{Optimal source distribution for the perturbed cellular
  flow~\eqref{eq:sf} with
  (a) \hbox{$\delta_1=0.05$}, \hbox{$\delta_2=0$},
  mixing enhancement factor~\hbox{$\cE=87.59$};
  (b) \hbox{$\delta_1=0$}, \hbox{$\delta_2=0.05$}, \hbox{$\cE=90.10$};
  (c) \hbox{$\delta_1=\delta_2=0.2$}, \hbox{$\cE=97.60$};
  (d) \hbox{$\delta_1=0.5$}, \hbox{$\delta_2=0$}, \hbox{$\cE=95.99$}.
  The perturbations all break the~$G_1$ symmetry and selects a linear
  combination of the degenerate eigenfunction in
  Fig.~\ref{fig:srcopt_cell}.  (See the caption to
  Fig.~\ref{fig:srcopt_pert} for a key to the background shading.)}
\label{fig:srcopt_cell_delta}
\end{figure}

Next, we resolve the degeneracy of the optimal source distribution by
setting~$\delta_1=0.05$ in~\eqref{eq:sf} while keeping~$\delta_2=0$, giving
the streamfunction shown as contour lines in the foreground of
Fig.~\ref{fig:srcopt_cell_delta1}.  The rolls now have a slight asymmetry that
breaks the rotational symmetry~$G_1$ of the unperturbed cellular flow.  As a
consequence, the normalized optimal eigenfunction is now unique, and is shown
as the shaded background in Fig.~\ref{fig:srcopt_cell_delta1}.  It is very
close to the degenerate eigenfunction on the right in
Fig.~\ref{fig:srcopt_cell}, and converges to it as~\hbox{$\delta_1\rightarrow
0$}.  The enhancement factor in this case is almost unchanged,~$87.59$.
Again, the cosine reference source has much higher enhancement factor
($86.18$, $1.6\%$ improvement for the optimal source) than the sine reference
source ($49.76$, $76.0\%$ improvement for the optimal source).

Finally, we set~$\delta_1=0$ and~$\delta_2=0.05$ in~\eqref{eq:sf}, with the
streamfunction shown as contours in the foreground of
Fig.~\ref{fig:srcopt_cell_delta2}.  As for the previous perturbation, this one
also breaks the~$G_1$ symmetry and causes the optimal eigenfunction to become
unique, but this time a superposition of the two degenerate eigenfunctions in
Fig.~\ref{fig:srcopt_cell} is selected.  The optimal enhancement
factor,~$90.10$, is again almost unchanged by this small perturbation.

To summarize this section, we presented three cases at fixed
diffusivity~$\kappa=0.01$.  The first was the cellular flow, for which we get
doubly-degenerate optimal eigenfunctions.  We then presented two
symmetry-breaking perturbations in turn, showing how these select a particular
mixture of the degenerate eigenfunctions to create a unique optimal source
distribution.  Since the perturbations are small, the optimal enhancement
factor in all these cases is about the same, showing an improvement of
about~$75$--$80\%$ over the reference source~$\sin x$, but only~$1$--$5\%$
over the reference source~$\cos x$.  This latter modest improvement is best
seen not as a failure of the optimization procedure, but as an advantage,
since robustness is always desirable.  In fact robustness can easily be gauged
by looking at the magnitude of the next largest eigenvalues, to see how far
they are from the dominant one(s).  Table~\ref{tab:eff} also shows that the
modest improvement over the cosine source is an accident, since for many other
velocity fields improvements well above~$50\%$ are seen for both sine and
cosine.  Two further examples for large perturbations are shown in
Figs.~\ref{fig:srcopt_cell_delta1b} and~\ref{fig:srcopt_cell_delta12}.

\subsection{Dependence on Diffusivity}
\label{sec:diffdep}

\begin{figure}
\psfrag{x}{{\tiny $x$}}
\psfrag{y}{{\tiny $y$}}
\psfrag{kappa = 0.1}{{\tiny $\kappa = 0.1$}}
\psfrag{kappa = 0.25}{{\tiny $\kappa = 0.25$}}
\psfrag{kappa = 0.5}{{\tiny $\kappa = 0.5$}}
\psfrag{kappa = 1}{{\tiny $\kappa = 1$}}
\psfrag{kappa = 5}{{\tiny $\kappa = 5$}}
\psfrag{kappa = 100}{{\tiny $\kappa = 100$}}
\begin{center}
\subfigure{\includegraphics[width=.3\textwidth]{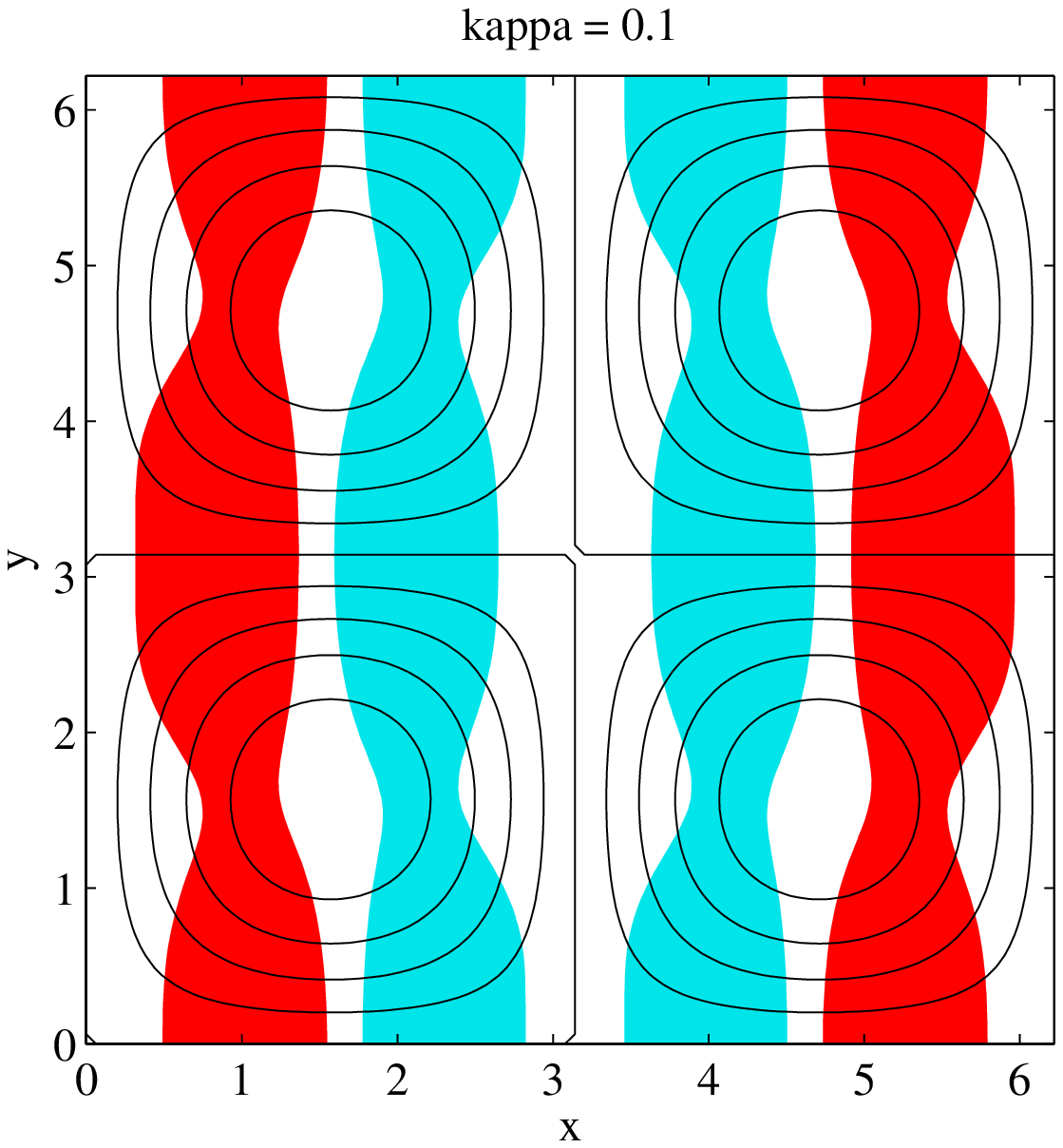}}
\hspace{.5em}
\subfigure{\includegraphics[width=.3\textwidth]{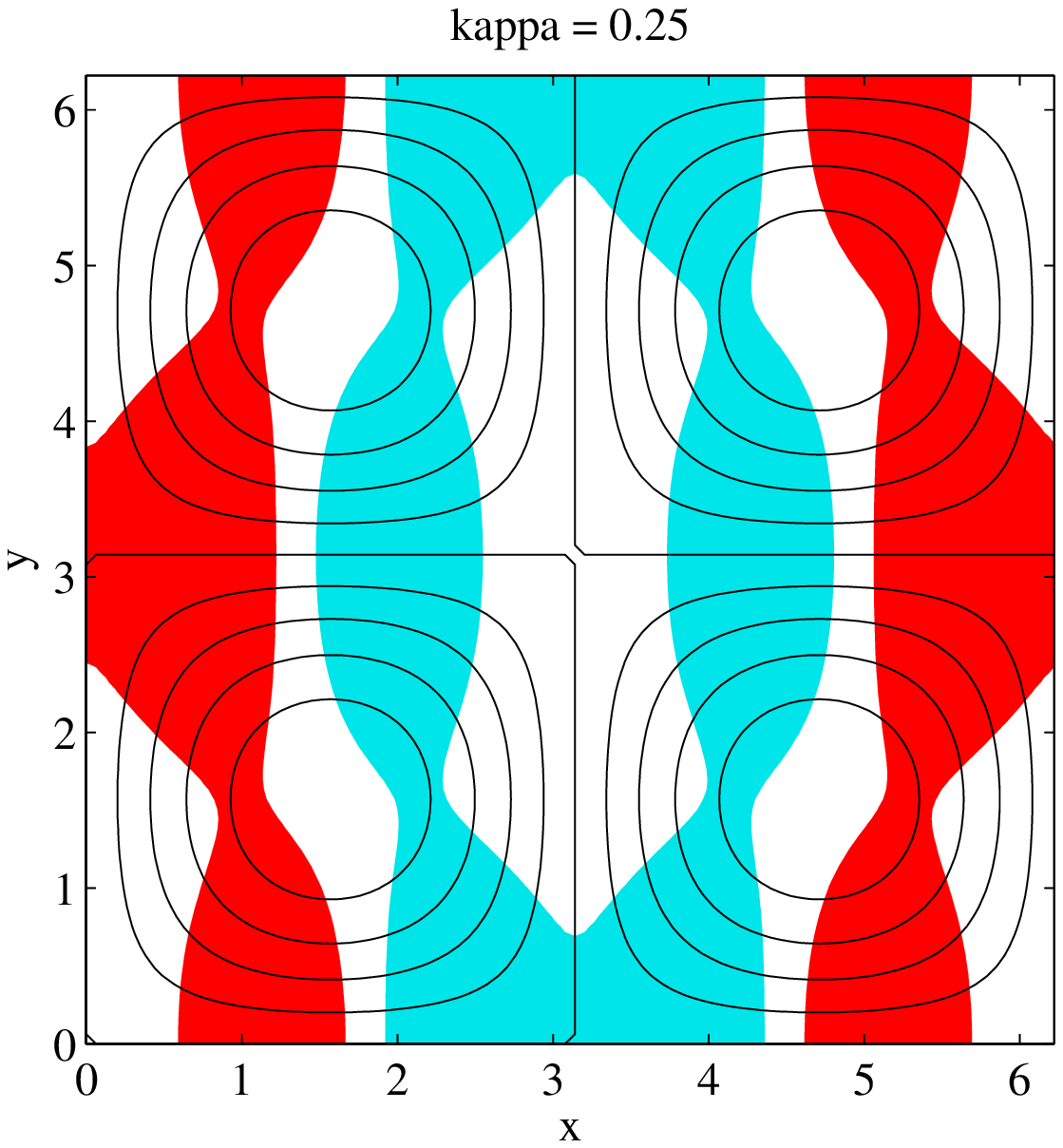}}
\hspace{.5em}
\subfigure{\includegraphics[width=.3\textwidth]{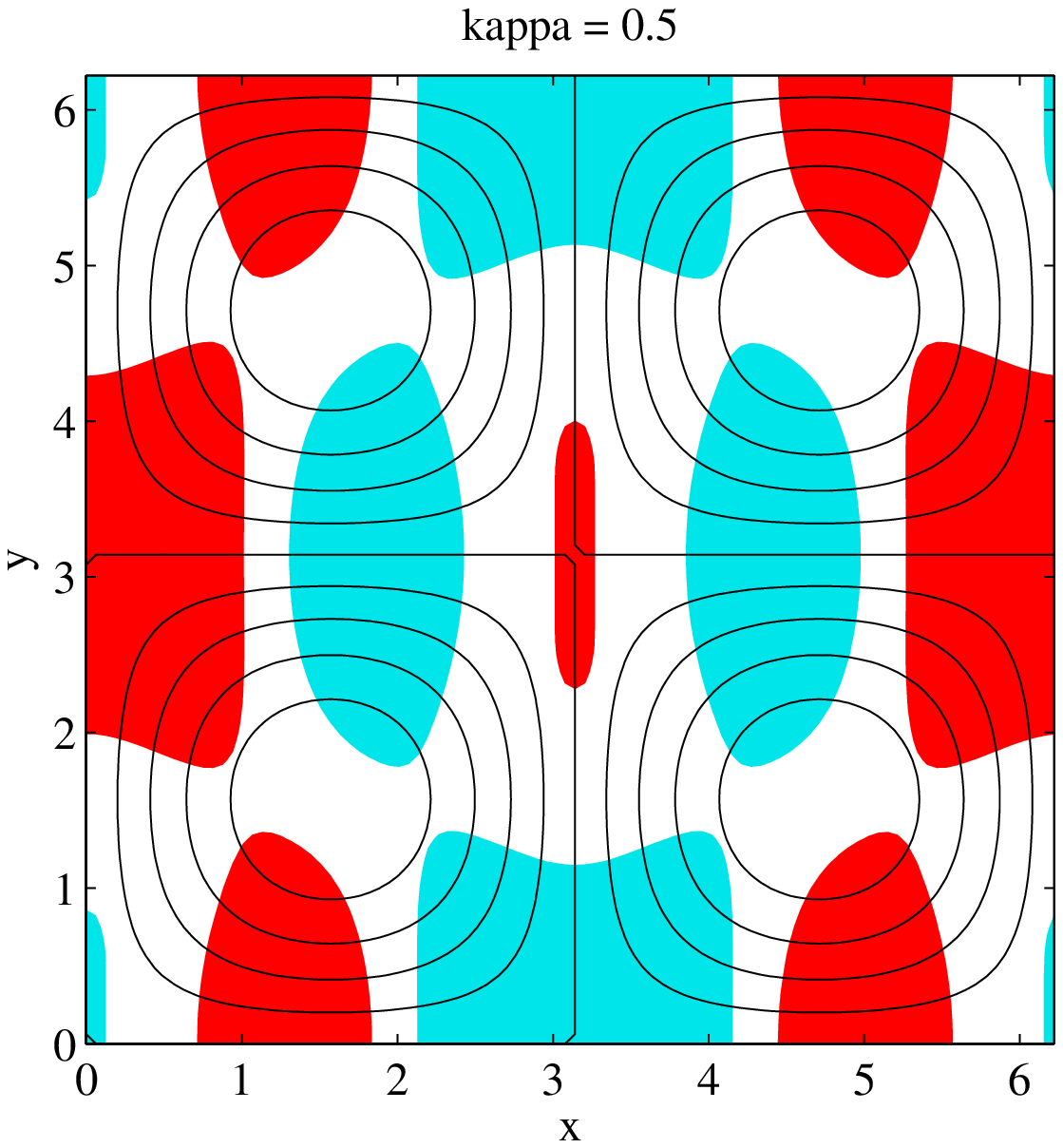}}

\subfigure{\includegraphics[width=.3\textwidth]{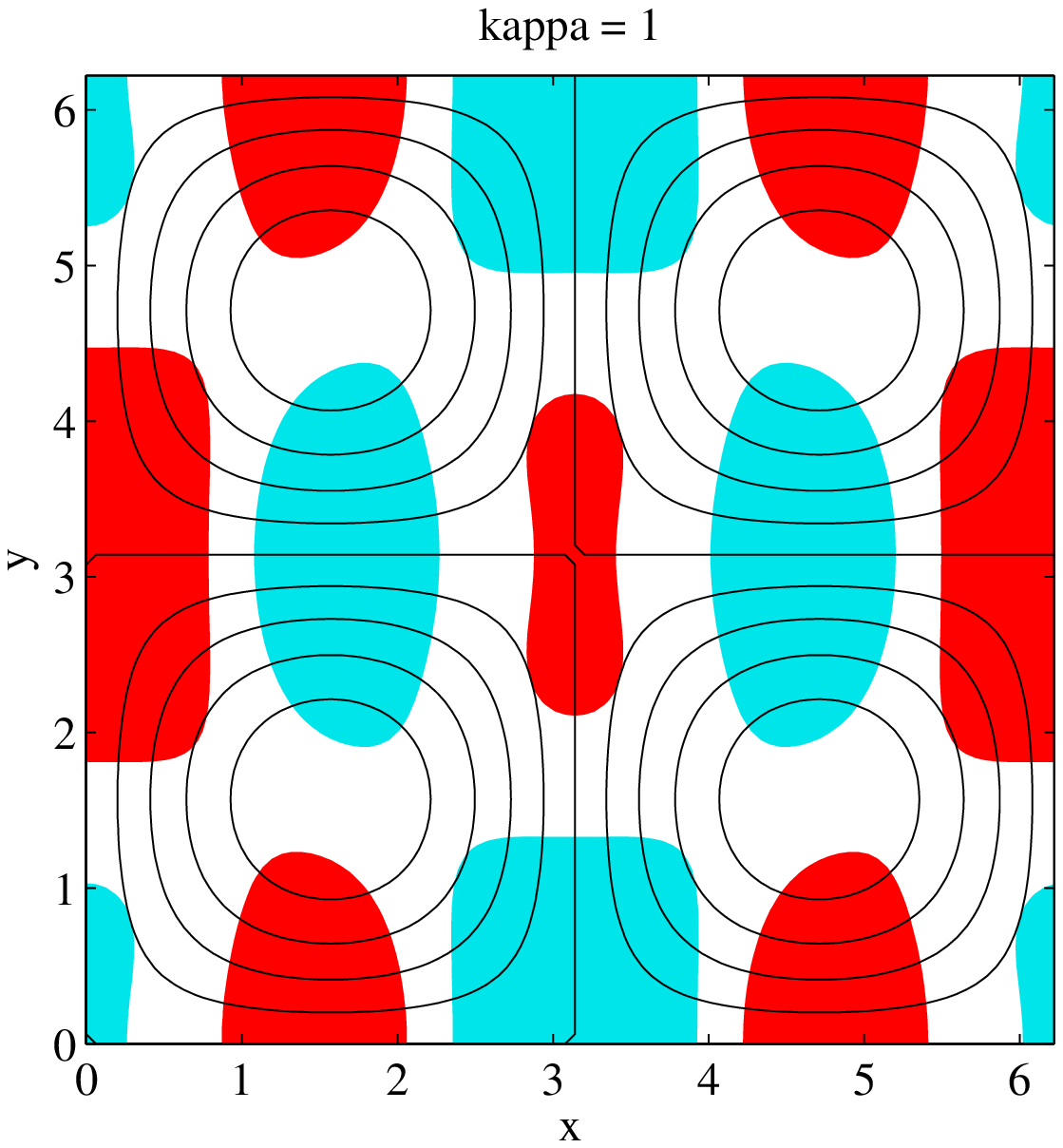}}
\hspace{.5em}
\subfigure{\includegraphics[width=.3\textwidth]{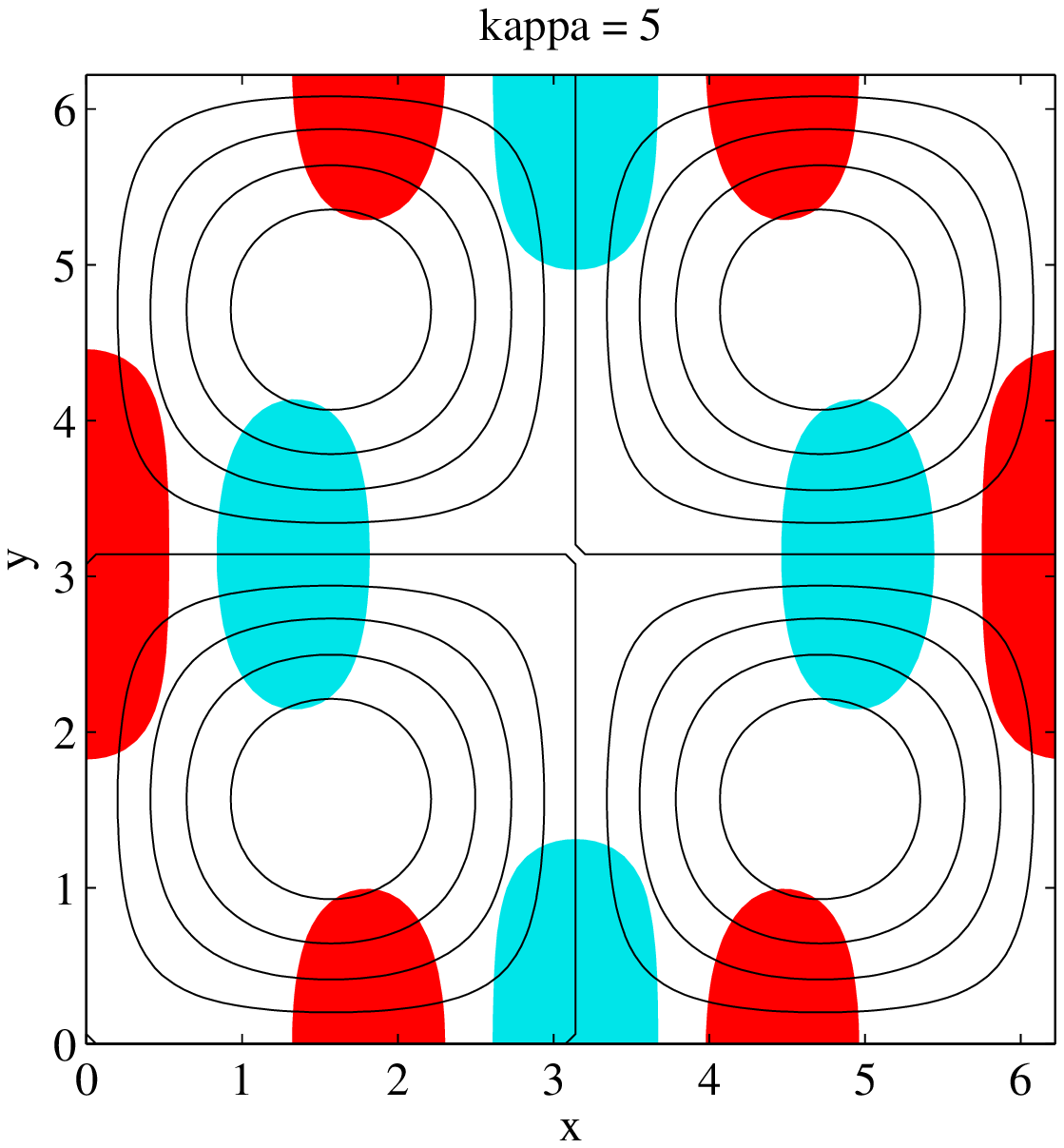}}
\hspace{.5em}
\subfigure{\includegraphics[width=.3\textwidth]{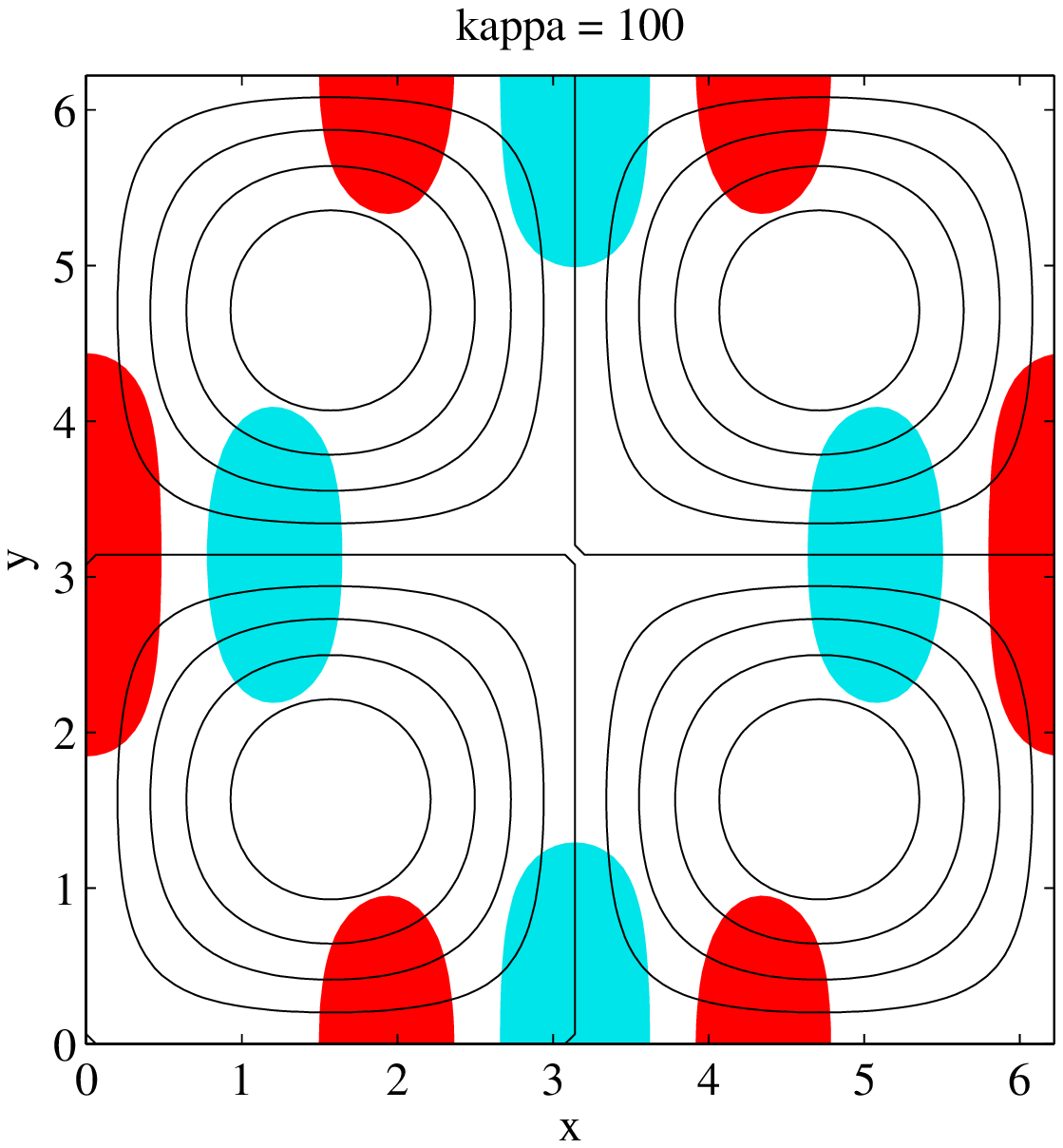}}
\end{center}
\caption{For the same flow as in Fig.~\ref{fig:srcopt_cell}, optimal source
  distribution for different values of the diffusivity~$\kappa$, increasing
  from top left to bottom right.  The eigenfunction is doubly-degenerate and
  corresponds to the one on the left in Fig.~\ref{fig:srcopt_cell}.  For both
  small and large~$\kappa$ the optimal source converges to an invariant
  eigenfunction.  In all cases there is no source of temperature over the
  elliptic stagnation points, but in the large~$\kappa$ case there are sources
  and sinks over some hyperbolic points.  (See the caption to
  Fig.~\ref{fig:srcopt_pert} for a key to the background shading.)}
\label{fig:srcopt_cell_kappa}
\end{figure}

Now we will fix~$\delta_1=\delta_2=0$ in~\eqref{eq:sf} and vary the
diffusivity,~$\kappa$.  For~$\kappa=0.01$, Fig.~\ref{fig:srcopt_cell}
shows the two degenerate optimal source eigenfunctions, and we will
follow the change in the one on the left as~$\kappa$ is varied.  In
Fig.~\ref{fig:srcopt_cell_kappa} we show the
change in the optimal source as~$\kappa$ is increased from~$0.1$
to~$100$. The optimal source distribution appears to become
independent of~$\kappa$ both for small~$\kappa$ and large~$\kappa$,
but the distributions are different.  The transition between the two
regimes occurs when~$\kappa$ is of order unity.  Though the two
asymptotic sources are very different, they respect the general
principles laid out in Section~\ref{sec:cellflow}: the source is
arranged for effective transport of hot onto cold and vice versa, and
regions of high speed are favored.  In particular, note that the
center of the rolls has a nearly zero, flat source distribution in all
cases.

Another perhaps surprising aspect of the large~$\kappa$ solution in
Fig.~\ref{fig:srcopt_cell_kappa} is that it has complicated structure.  In
this large diffusivity limit, one would expect diffusion to dominate and
gradients to be smoothed out.  But since our mixing enhancement
factor~\eqref{eq:mixeff} compares the variance to the unstirred case, which
already has very low variance, any amount of improvement will count.  Hence,
the complicated source for large~$\kappa$ in Fig.~\ref{fig:srcopt_cell_kappa}
only gives a minute improvement to the enhancement factor.  The large~$\kappa$
optimal solution is particular in that it has some hot and cold spots
localized over hyperbolic stagnation points.  This is probably due to the high
speeds along the separatrices being favored, even at the cost of straddling
hyperbolic stagnation points a little.

\begin{figure}
\psfrag{E-1}{$\cE-1$}
\psfrag{kappa}{$\kappa$}
\psfrag{k-1}{{\tiny $\kappa^{-1}$}}
\psfrag{k-2}{{\tiny $\kappa^{-2}$}}
\psfrag{sin}{$\sin$}
\psfrag{cos}{$\cos$}
\begin{center}
\includegraphics[width=.6\textwidth]{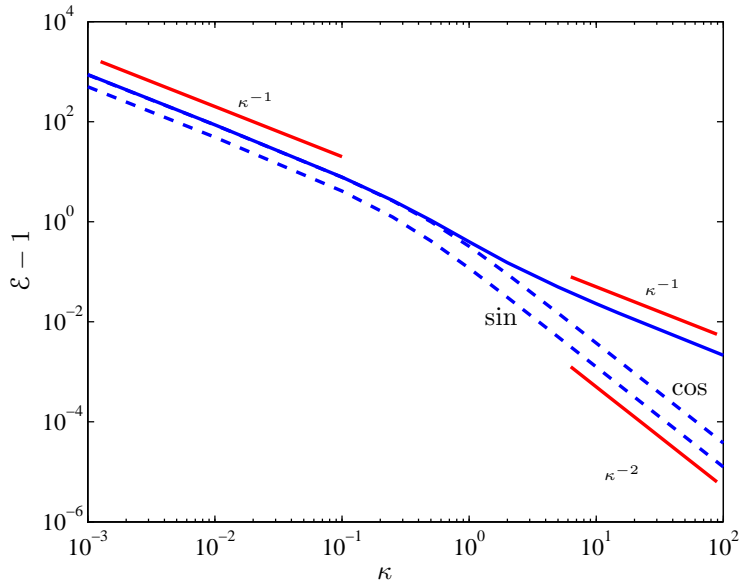}
\end{center}
\caption{For the flow with streamfunction as in Fig.~\ref{fig:srcopt_cell},
  mixing enhancement factor~$\cE-1$ as a function of the diffusivity~$\kappa$:
  optimal source (solid line), and~$\sin x$ and~$\cos x$ reference sources
  (dashed lines).  For small~$\kappa$, the enhancement factor scales
  like~$\kappa^{-1}$.  For large~$\kappa$, the optimal solution returns to
  a~$\kappa^{-1}$ approach to unity after a brief dip, while the reference
  source solution approaches unity as~$\kappa^{-2}$.}
\label{fig:effplot_cell_kappa}
\end{figure}

Figure~\ref{fig:effplot_cell_kappa} shows how the mixing enhancement factor
varies as a function of the diffusivity.
The solid line is for the optimal source, the dashed lines for the reference
sources~$\sin x$ and~$\cos x$.  For small~$\kappa$, the enhancement factor of
all sources scales as~$\kappa^{-1}$: this is the `classical' scaling discussed
in~\cite{Thiffeault2004,DoeringThiffeault2006,Shaw2007}, where the enhancement
factor is linear in the P\'eclet number.  It has been rigorously proved
in~\cite{Thiffeault2004,DoeringThiffeault2006,Shaw2007} that this scaling is
optimal over all possible sources and velocity fields.

For~$\kappa$ near unity, the optimal enhancement factor has a small dip before
converging towards unity as~$\kappa^{-1}$ for large~$\kappa$.  In
contrast, the reference enhancement factors converges to unity
as~$\kappa^{-2}$.  This last scaling holds for the uniform flow of
Section~\ref{sec:perturb} when expanded in large~$\kappa$, and is
verified for other flows and sources as
well~\cite{Thiffeault2004,DoeringThiffeault2006,Shaw2007}.

In summary, the optimal source distribution becomes independent of~$\kappa$
for both large and small $\kappa$, but of course for large~$\kappa$ the
efficiency gain is minimal (since the~$\Ltwo$--norm of the velocity is fixed).
For small $\kappa$ the efficiency gain is a constant multiple of the reference
sources, but this multiple is fairly small for~$\cos x$ ($1.01$), showing that
optimization is very robust but not necessarily always worthwhile.  Overall,
the optimal enhancement factor scales as~$\kappa^{-1}$, with a momentary break
in the scaling that corresponds to the complicated change in topology seen in
Fig.~\ref{fig:srcopt_cell_kappa} for $\kappa$ near unity.

\subsection{Dependence on Exponent $\pexp$}
\label{sec:pexpdep}

Our final study will be to examine the behavior of the optimal enhancement
factor as~$\pexp$ is varied in~\eqref{eq:mixeff}.  In
Sections~\ref{sec:cellflow}--\ref{sec:diffdep} we used~$\pexp=0$; now we
fix~$\delta_1=\delta_2=0$, $\kappa=0.01$, and allow~$\pexp$ to vary over
negative and positive values.
\begin{figure}
\psfrag{x}{{\tiny $x$}}
\psfrag{y}{{\tiny $y$}}
\psfrag{p = -1}{{\tiny $\pexp = -1$}}
\psfrag{p = -0.25}{{\tiny $\pexp = -0.25$}}
\psfrag{p = 0}{{\tiny $\pexp = 0$}}
\psfrag{p = 0.5}{{\tiny $\pexp = 0.5$}}
\psfrag{p = 1}{{\tiny $\pexp = 1$}}
\psfrag{p = 2}{{\tiny $\pexp = 2$}}
\begin{center}
\subfigure{\includegraphics[width=.3\textwidth]{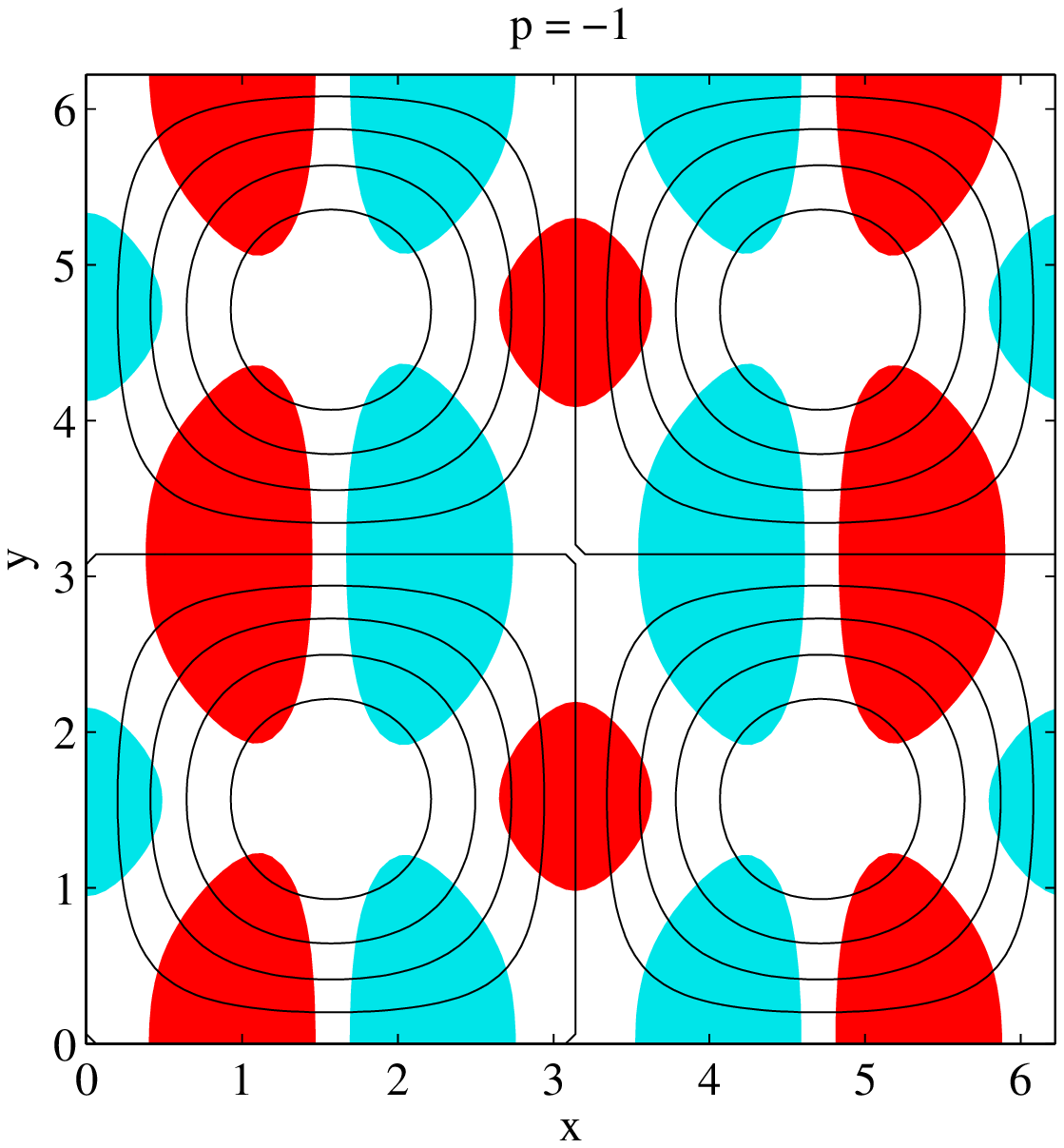}}
\hspace{.5em}
\subfigure{\includegraphics[width=.3\textwidth]{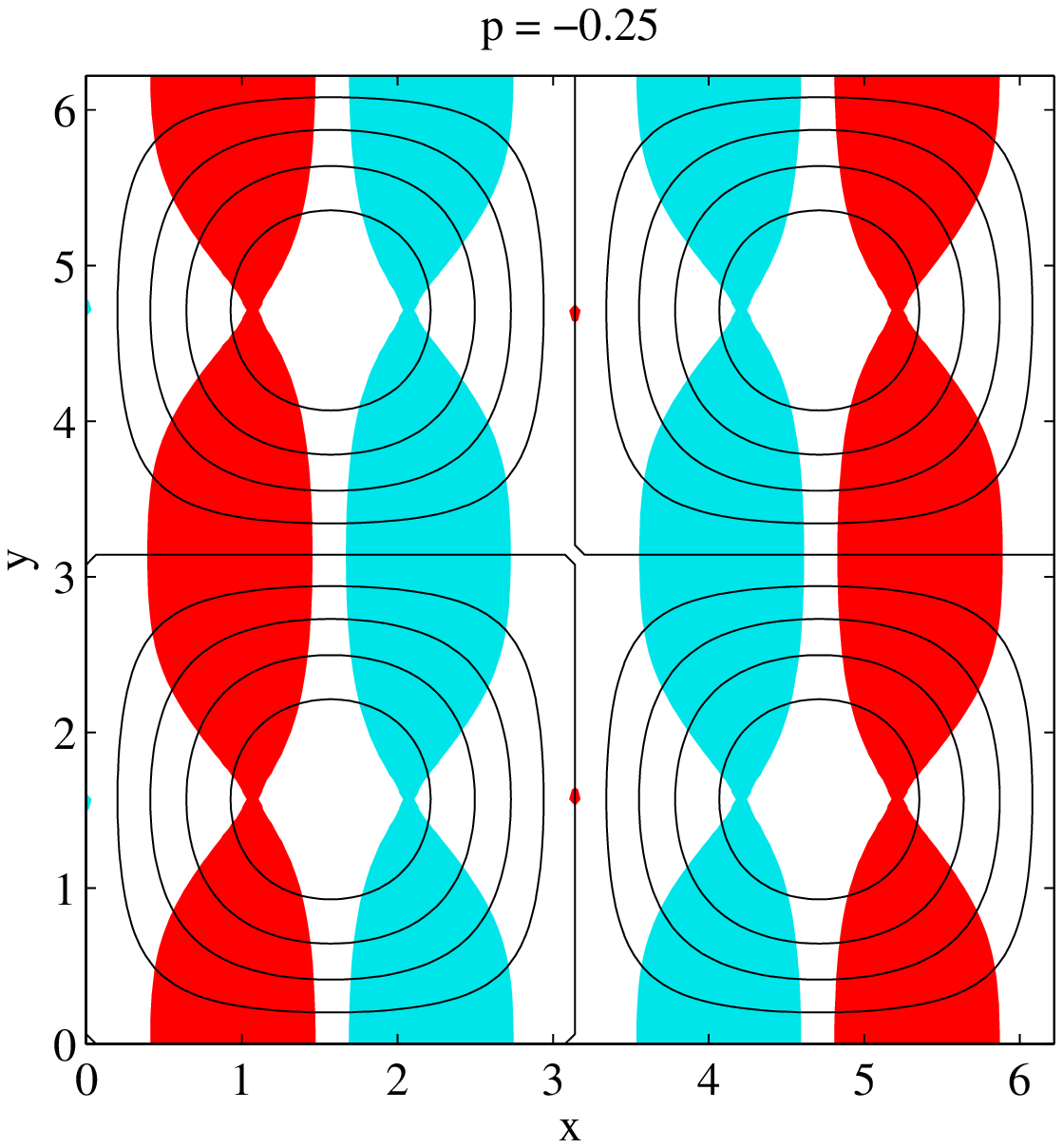}}
\hspace{.5em}
\subfigure{\includegraphics[width=.3\textwidth]{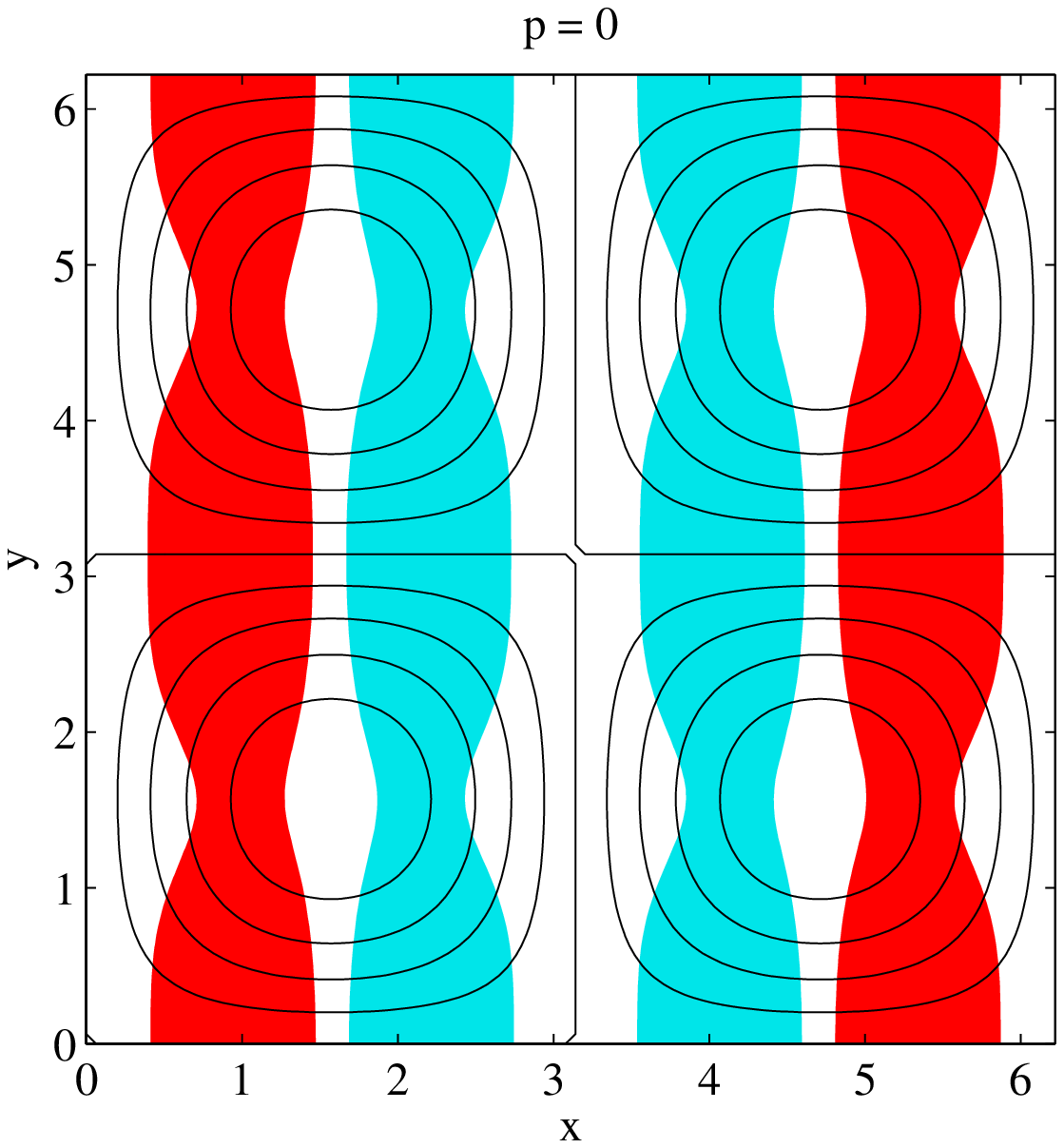}}

\subfigure{\includegraphics[width=.3\textwidth]{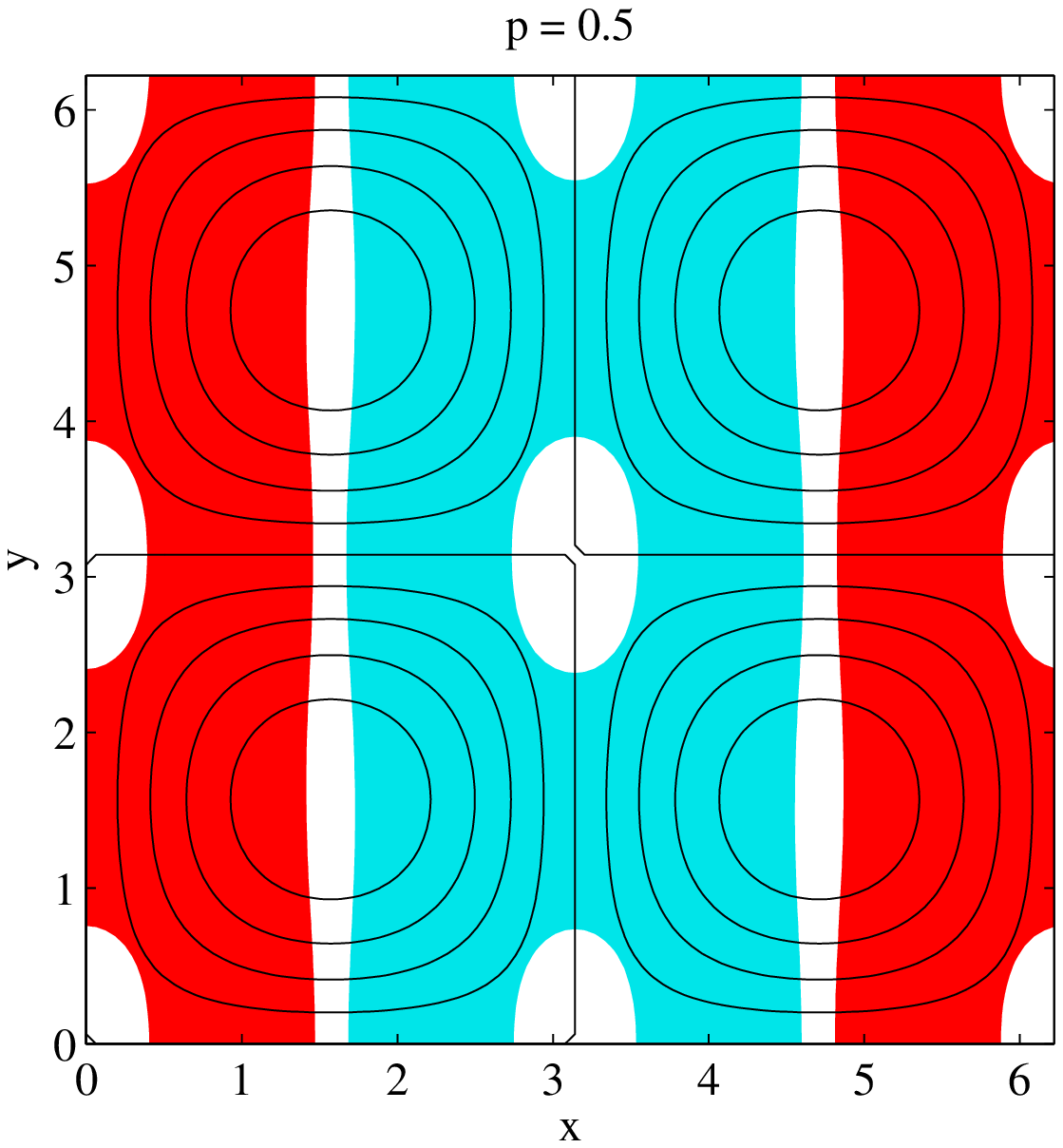}}
\hspace{.5em}
\subfigure{\includegraphics[width=.3\textwidth]{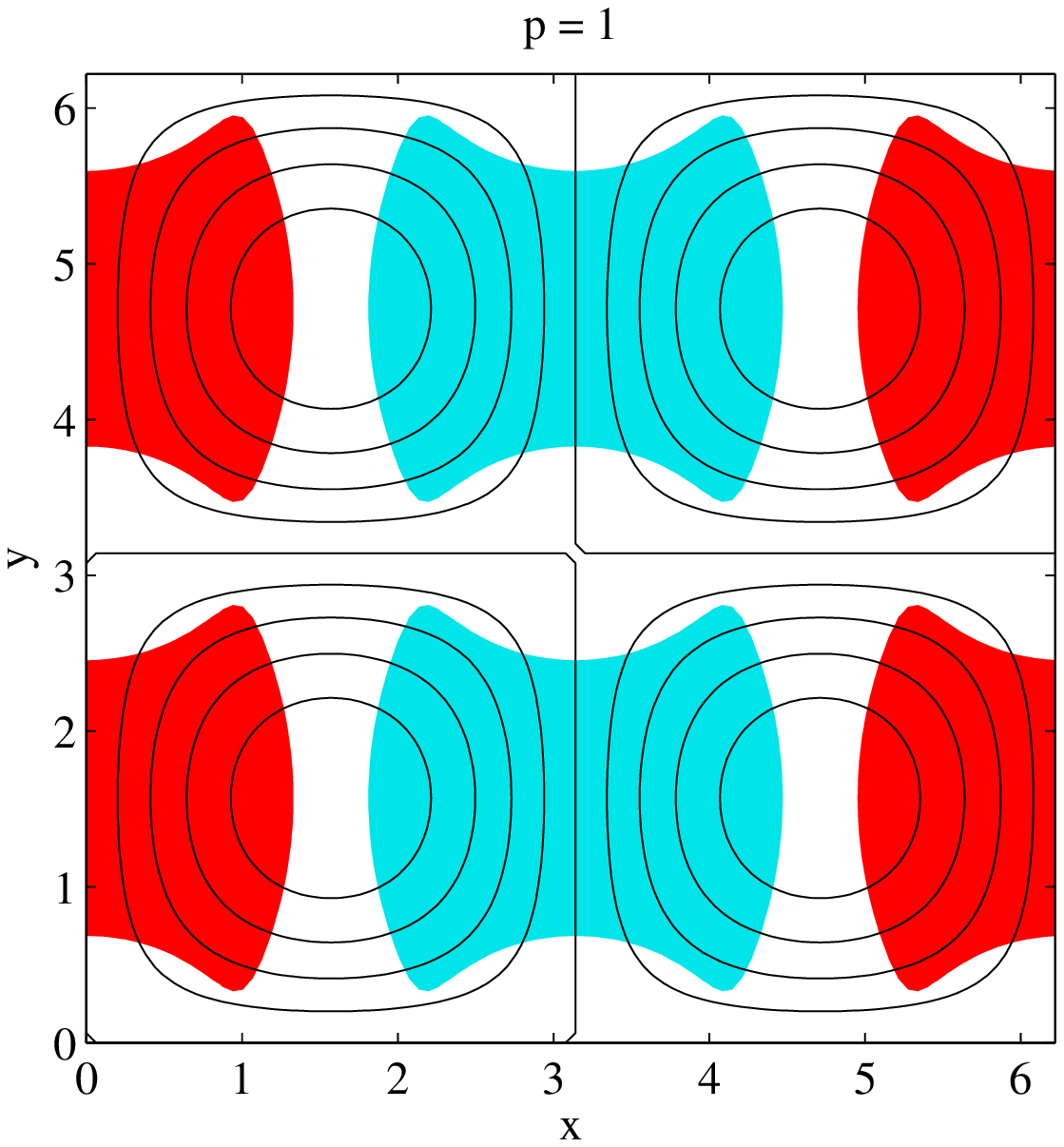}}
\hspace{.5em}
\subfigure{\includegraphics[width=.3\textwidth]{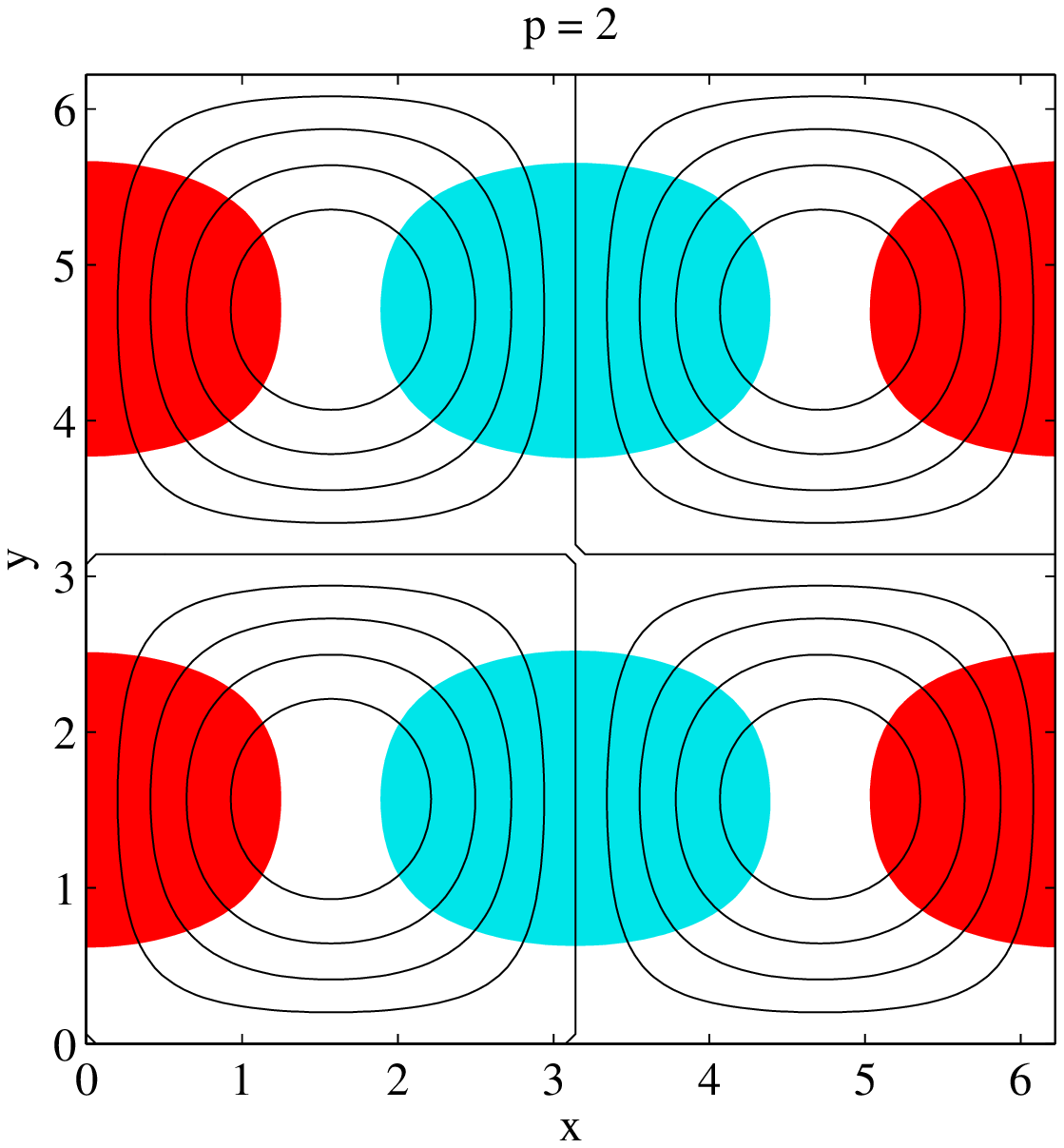}}
\end{center}
\caption{For the same flow as in Fig.~\ref{fig:srcopt_cell}, optimal source
  distribution for different exponents~$\pexp$, increasing from top left to
  bottom right.  The eigenfunction is doubly-degenerate and corresponds to the
  one on the left in Fig.~\ref{fig:srcopt_cell}.  For both small and
  large~$\pexp$ the optimal source converges to an invariant eigenfunction.
  In all cases there are no sources or sinks of temperature over the
  stagnation points.  (See the caption to Fig.~\ref{fig:srcopt_pert} for a key
  to the background shading.)}
\label{fig:srcopt_cell_p}
\end{figure}
Figure~\ref{fig:srcopt_cell_p} shows the optimal source distributions
for~$\pexp$ varying from~$-1$ to~$2$.  For both negative and
positive~$\pexp$, the optimal source distribution converges rapidly to
invariant patterns, and the two extremes in
Fig.~\ref{fig:srcopt_cell_p} are representative of those asymptotic
patterns.  The situation is thus entirely analogous to the case where
diffusivity was varied (Fig.~\ref{fig:srcopt_cell_kappa}).

The top-left picture in Fig.~\ref{fig:srcopt_cell_p} (negative~$\pexp$) shows
small, localized sources and sinks.  In contrast, the bottom-right picture in
Fig.~\ref{fig:srcopt_cell_p} (positive~$\pexp$) shows large, regular localized
sources and sinks.  In fact, what is striking about the pattern is its
simplicity: it is what one might take as a guess at an efficient source
distribution, with no added frills.  Thus, a high power of~$\pexp$ might be
useful in situations where a simple configuration is necessary due to
engineering constraints.  The reason for the simplicity is that spatial
variations in the source favor the diffusion operator in~$\LL$, and
as~$\pexp\rightarrow\infty$ these are magnified.  Thus, the source must remain
as spatially simple as possible while trying to maximize alignment with the
velocity.  As~$\pexp\rightarrow-\infty$, spatial variations of the source are
downplayed by the norm, allowing more complexity.

\begin{figure}
\psfrag{E-1}{$\cE_\pexp-1$}
\psfrag{2.2-p}{$(2.2)^{-\pexp}$}
\psfrag{2.2p}{$(2.2)^{\pexp}$}
\psfrag{p}{\pexp}
\psfrag{sin}{$\sin$}
\psfrag{cos}{$\cos$}
\begin{center}
\includegraphics[width=.6\textwidth]{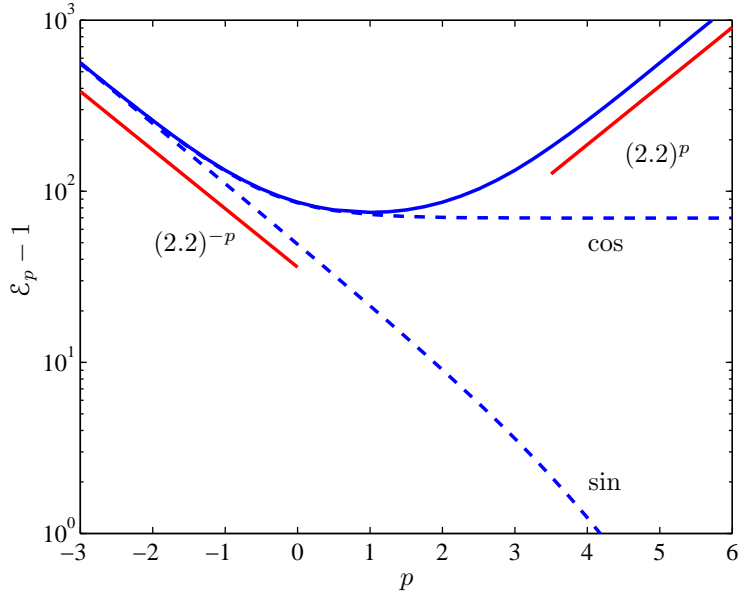}
\end{center}
\caption{For the flow with streamfunction as in
  Fig.~\ref{fig:srcopt_cell}, mixing enhancement factor~$\cE_\pexp-1$ as a
  function of the exponent~$\pexp$: optimal source (solid line),
  and~$\sin x$ and~$\cos x$ reference sources (dashed line).  The
  optimal enhancement factor is symmetric about~\hbox{$\pexp=1$}, and
  for~$\lvert\pexp\rvert\gg 1$ it grows
  as~$(2.2)^{\lvert\pexp\rvert}$.}
\label{fig:effplot_cell_p}
\end{figure}

Figure~\ref{fig:effplot_cell_p} shows how the optimal mixing enhancement
factor
varies as a function of~$\pexp$.  For~$\pexp\rightarrow-\infty$, the
enhancement factor goes to infinity, as does the enhancement factor of the two
reference sources.  For~$\pexp\rightarrow\infty$, the enhancement factor also
goes to infinity, but the reference source enhancement factors now approach
constants (the constant is~$1$ for~$\sin x$).  Again, we are seeing the effect
of the diffusion term dominating when gradients are present, since these are
amplified by~$\mlapl^{\pexp/2}$.  Note that the curve is symmetric
about~\hbox{$\pexp=1$}, which leads to a minimum there: whether this is true
in general has not been determined, but we have not found a counterexample.
In Appendix~\ref{sec:psym} we provide a partial proof by explicitly finding
the symmetry between the operators~$\cA_{2-\pexp}$ and~$\cA_\pexp$, but only
for large~$\kappa$.

\section{Discussion}
\label{sec:discussion}

In both the perturbation problem (Section~\ref{sec:perturb}) and the
numerical examples (Section~\ref{sec:numerics}), the optimal source
distributions tend to exhibit the following features:
\begin{enumerate}
\item Avoidance of stagnation points of the flow, especially of
  elliptic type;
\item Localization over regions of rapid flow;
\item Alignment of the source contours perpendicular to the local
  velocity, so that hot is swept onto cold and vice versa.
\end{enumerate}
For a shear flow perturbation (Section~\ref{sec:shearwavy}), the optimal
source bulges out over regions of faster flow.  In contrast, a wavy flow
perturbation leaves the optimal source unchanged over from a uniform flow,
suggesting that localization is a more important effect than alignment.  This
also demonstrates that the optimal solution for source optimization does not
necessarily correspond to the optimal solution for velocity optimization,
given that optimal source.  This is because at fixed energy the wavy flow
always decreases the optimal enhancement factor from that of the uniform flow,
for the same source distribution.

These considerations show that the mixing enhancement factor and optimization
procedure described in this paper (Section~\ref{sec:optimization}) behave in a
natural manner.  Furthermore, the procedure yields a global maximum
(Section~\ref{sec:maximal}), is numerically well-behaved, and is easy to
implement.  Indeed, a three-dimensional application to a real system is well
within the realm of feasibility: the examples we presented here (except for
the uniform flow) were two-dimensional, but for smooth flows the sparse nature
of the advection-diffusion operator in Fourier space means that large problems
can be solved with a modest amount of computer power and memory.

The optimal source distribution becomes independent of the
diffusivity~$\kappa$ for both large and small~$\kappa$
(Section~\ref{sec:diffdep}), and for exponent~$\pexp$ [see
Eq.~\eqref{eq:mixeff}] negative and large or positive and large
(Section~\ref{sec:pexpdep}).  None of the source distributions achieved in
these cases resemble each other.  However, we observed that the eigenvalue
spectrum of~\eqref{e:eu_la_p_GE} is always symmetric about~$\pexp=1$ (and has
a minimum there), which implies that the operators (and thus the optimal
source distributions) are related by a unitary transformation we were unable
to find in general (but see Appendix~\ref{sec:psym} for a partial result).
The large positive~$\pexp$ case is particularly interesting, since it favors
source distributions that are very smooth, a desirable feature in practical
applications.  Another attractive feature we observed is the robustness of the
optimal solution in the cases considered (Section~\ref{sec:cellflow}), but
this will not necessarily hold in general.

To widen the applicability of the procedure, a few complications will have to
be introduced.  First, time-dependence of the flow and the source is
desirable, which will make the variational problem more difficult to solve in
principle.  Second, and more importantly, there remains the much more
difficult problem of optimizing the velocity field given a source
distribution.  The variational problem is easily formulated, but does not
present itself in the nice generalized eigenvalue problem
[Eq.~\eqref{e:eu_la_p_GE}] we saw here.  The third and ultimate goal is a full
dynamical coupling to the Navier--Stokes equation, with buoyancy and other
effects included as appropriate.

Throughout the paper we spoke of `mixing' because our description
involves the interplay of stirring and molecular diffusion.  However,
we highlighted the more interesting limit of small
diffusivity~$\kappa$ (large P\'eclet), since this is the situation in
which stirring is more pertinent.  In that case, it is clear that the
pushing of hot fluid onto cold regions and vice versa is better
described as `transport' rather than `mixing'.  Thus, perhaps in that
limit we should speak of a `transport enhancement factor', but since our
analysis applies even for very large~$\kappa$ we have kept the
terminology, in spite of the fact that the velocity fields presented
here do not lead to creation of small scales and thus may not `mix'
very well by some criteria.

\jltnote{Do antieffective sources exist?  Haven't found any
  eigenvalues less that unity that persisted when more modes were
  added.}

\jltnote{The unnormalized enhancement factor leads to small-scale instability.
  Is that right?  For~$\pexp$ large enough, get same answer whether normalized
  by diffusive solution or by source.}

\begin{acknowledgments}
  The authors thank Charles R. Doering for inspiring discussions.
  J.-L.T. was supported in part by the UK Engineering and Physical
  Sciences Research Council grant GR/S72931/01.
\end{acknowledgments}


\appendix

\section{Symmetry of Optimal Mixing Enhancement Factor About~$\pexp=1$}
\label{sec:psym}

\mathnotation{\cB}{\mathcal{B}}
\mathnotation{\cP}{\mathcal{P}}

In this appendix we will motivate the symmetry of the optimal
enhancement factor curve about~$\pexp=1$, as evident in
Fig.~\ref{fig:effplot_cell_p}.  We use the self-adjoint
form~\eqref{eq:ge}: let
\begin{equation*}
  \cB_\pexp[\uv] \ldef \cAz_\pexp^{-1/2}\cA_\pexp[\uv]
  \,\cAz_\pexp^{-1/2}\,,
\end{equation*}
where we have explicitly shown the dependence of the operators
on~$\uv(\xv)$.  We show below that
\begin{equation}
  \cP\,\cB_{2-\pexp}[\uv]\,\cP = \cB_\pexp[\cP\uv]
  + \Order{\kappa^{-2}}\,,
  \label{eq:Psym}
\end{equation}
where~$\cP$ is the parity change operator, defined by~$\cP\,\rsrc(\xv)
= \rsrc(-\xv)$ and~$\cP\uv(\xv) = -\uv(-\xv)$, with~$\cP^{-1} = \cP$.
Since the spectrum is unchanged by the
substitution~$\uv(\xv) \rightarrow -\uv(-\xv)$,
establishing~\eqref{eq:Psym} proves that~$\cB_{2-\pexp}[\uv]$
and~$\cB_\pexp[\uv]$ have the same spectrum, which explains the
symmetry of the enhancement factor about~$\pexp=1$.  The eigenfunctions are
related by a parity change.  Unfortunately, Eq.~\eqref{eq:Psym} is
only an asymptotic result valid for large~$\kappa$, and we do not know
the general form of the symmetry~$\cP$ for smaller~$\kappa$, though
numerical evidence suggests it exists.

To show~\eqref{eq:Psym} directly, we expand
\begin{equation*}
  \cP\,\cB_{2-\pexp}[\uv(\xv)]\,\cP = 1 + \kappa^{-1}
  \bigl(\mlapl^{-\tfrac{\pexp}{2}}\,\cP\,
  \uv(\xv)\cdot\grad\cP\,\mlapl^{\tfrac{\pexp}{2}-1}
  - \mlapl^{\tfrac{\pexp}{2}-1}\,\cP\,
  \uv(\xv)\cdot\grad\cP\,\mlapl^{-\tfrac{\pexp}{2}}
  \bigr) + \Order{\kappa^{-2}}
\end{equation*}
where we used the commutativity of~$\lapl$ and~$\cP$.  We also have
\begin{equation*}
  \cB_\pexp[-\uv(-\xv)] = 1 + \kappa^{-1}
  \bigl(\mlapl^{-\tfrac{\pexp}{2}}\,
  \uv(-\xv)\cdot\grad\,\mlapl^{\tfrac{\pexp}{2}-1}
  - \mlapl^{\tfrac{\pexp}{2}-1}\,
  \uv(-\xv)\cdot\grad\,\mlapl^{-\tfrac{\pexp}{2}}
  \bigr) + \Order{\kappa^{-2}}.
\end{equation*}
Now Eq.~\eqref{eq:Psym} follows from~$\cP\,\uv(\xv)\cdot\grad\cP =
\uv(-\xv)\cdot\grad$, since both~$\uv$ and~$\grad$ reverse direction
under parity change.

\end{document}